\documentclass[twocolumn]{aastex7}
\usepackage{amsmath}

\DeclareMathAlphabet\mathbfcal{OMS}{cmsy}{b}{n}
\begin{document}

\title{A \textsl{Spitzer Space Telescope} Exploration Science Program to Search for Y Dwarf Variability}

\author[orcid=0000-0001-7780-3352]{Michael C. Cushing}
\affiliation{Ritter Astrophysical Research Center, Department of Physics and Astronomy, University of Toledo, Toledo, OH 43606, USA}
\email[show]{michael.cushing@utoledo.edu}

\author[orcid=0000-0001-5886-1354]{Jesica L. Trucks}
\affiliation{Ritter Astrophysical Research Center, Department of Physics and Astronomy, University of Toledo, Toledo, OH 43606, USA}
\email{jmactrucks@gmail.com}

\author[orcid=0000-0003-3702-0382]{Kevin K. Hardegree-Ullman}
\affiliation{IPAC, Mail Code 100-22, California Institute of Technology, 1200 E. California Boulevard, Pasadena, CA 91125, USA}
\email{kevinkhu@caltech.edu}

\author[orcid=0000-0002-6523-9536]{Adam J. Burgasser}
\affiliation{Center for Astrophysics and Space Science, University of California San Diego, La Jolla, CA 92093, USA}
\email{aburgasser@ucsd.edu}

\author[orcid=0000-0002-0221-6871]{Sean J. Carey}
\affiliation{IPAC, Mail Code 100-22, California Institute of Technology, 1200 E. California Boulevard, Pasadena, CA 91125, USA}
\email{cary@ipac.caltech.edu}

\author[orcid=0000-0002-9843-4354]{Jonathan J. Fortney}
\affiliation{Department of Astronomy and Astrophysics, University of California, Santa Cruz, CA 95064, USA}
\email{jfortney@ucolick.org}

\author[orcid=0000-0001-5072-4574]{Christopher R. Gelino}
\affiliation{IPAC, Mail Code 100-22, California Institute of Technology, 1200 E. California Boulevard, Pasadena, CA 91125, USA}
\email{cgelino@ipac.caltech.edu}

\author[orcid=0000-0002-8916-1972]{John E. Gizis}
\affiliation{Department of Physics and Astronomy, University of Delaware, Newark, DE 19716, USA}
\email{gizis@udel.edu}

\author[orcid=0000-0003-4269-260X]{J. Davy Kirkpatrick}
\affiliation{IPAC, Mail Code 100-22, California Institute of Technology, 1200 E. California Boulevard, Pasadena, CA 91125, USA}
\email{davy@ipac.caltech.edu}

\author[orcid=0000-0002-3681-2989]{Sandy Leggett}
\affiliation{Gemini Observatory/NSF’s NOIRLab, 670 N. A’ohoku Place, Hilo, HI 96720, USA}
\email{sleggett@gemini.edu}

\author[orcid=0000-0001-7875-6391]{Gregory N. Mace}
\affiliation{Department of Astronomy, The University of Texas, Austin, TX 78712, USA}
\email{gmace@astro.as.utexas.edu}

\author[orcid=0000-0002-5251-2943]{Mark S. Marley}
\affil{Lunar \& Planetary Laboratory, University of Arizona, Tucson, AZ 85721,USA}
\email{marksmarley@arizona.edu}

\author[orcid=0000-0002-4404-0456]{Caroline V. Morley}
\affiliation{Department of Astronomy, The University of Texas, Austin, TX 78712, USA}
\email{cmorley@utexas.edu}

\newcommand{\cho}{[3.6]}
\newcommand{\cht}{[4.5]}

%\collaboration{all}{The Terra Mater collaboration}

\begin{abstract}

We present the results of a \textsl{Spitzer Space Telescope} Exploration Science Program to search for and characterize variability in Y dwarfs.  We observed 14 Y dwarfs over a 24 hr period at [3.6] and [4.5] and then repeated the observations a few months later.  We add two Y dwarfs, WD 0806$-$661B and WISE J085510.83$-$071442.5, that were also observed with \textsl{Spitzer} so that our sample includes all Y dwarfs observed for variability with \textsl{Spitzer}.  We infer variability fractions of 59\%$\pm$15\% and  64$^{+10\%}_{-13\%}$ for \cho\, and \cht\, in the first epoch, and 35\%$^{+17\%}_{-11\%}$ and 75\%$^{+8\%}_{-15\%}$ in the second epoch.  We also find that Y dwarf \textsl{Spitzer} light curves are generally stable over timescales of months, but in some cases can show clear changes in amplitude.   Combining our results with a similar \textsl{Spitzer} survey of L and T dwarfs by \citeauthor{2015ApJ...799..154M}, we find the mid-infrared variability fraction of L, T, and Y dwarfs weakly supports the hypothesis that brown dwarf variability is caused by variations in the horizonal and/or vertical structure of condensate clouds.  

\end{abstract}



\section{Introduction}

%\textcolor{red}{quasi periodic function? Periodogram including errors, Kepler, python.}

%\textcolor{red}{Must say inclination effects variability.}

The atmospheres of the coolest stars and brown dwarfs are distinct from those of hotter stars because the rate at which they release energy to space is controlled primarily by molecular and condensate opacity sources rather than atomic and continuum opacity sources \citep[e.g.,][]{2014PASA...31...43B}.  As a result, the emergent spectra of these cool objects are sculpted by broad absorption bands of TiO, H$_2$O, CH$_4$, and NH$_3$ and the relatively smooth opacities of condensates like enstatite (MgSiO$_3$), forsterite (MgSi$_2$O$_4$), and water ice \citep{2000ASPC..212..152M}.  It is the waxing and waning of these molecular absorption bands with decreasing effective temperature ($T_\mathrm{eff}$) that give rise to the MLTY spectral sequence \citep[e.g.,][]{2005ARA&A..43..195K,2009ssc..book.....G,2014ASSL..401..113C}.

Based on our experience with the giant planets in our solar system, the scores of condensate species that form in the atmospheres of these so-called ultracool dwarfs (with spectral types M7 and later) are believed to gravitationally settle and form clouds. Indeed model atmospheres that include simple cloud models provide a reasonably good match to the observed spectral energy distribution of the L and T dwarfs \citep{2005ApJ...621.1033T,2008ApJ...678.1372C,2009ApJ...702..154S,2011A&A...529A..44W}\footnote{\cite{2015ApJ...804L..17T} and \cite{2016ApJ...817L..19T} alternatively suggested that condensate clouds are not required in order to match the observations of the LTY dwarfs.}.  These models generally assume a uniform cloud coverage but the heterogeneous nature of the clouds on our giant planets \citep[e.g.,][]{1969ApJ...157L..63W} belies this simplifying assumption.  With heterogeneous clouds comes the possibility that the integrated light intensities of ultracool dwarfs may vary due to the modulation of the cloud structures by the rotation period of the object \citep{2001ApJ...556..872A} and/or changes in the clouds structures themselves\footnote{Magnetic activity can also result in variability for the hottest ultracool dwarfs but it is generally believed that the atmospheres of the cooler brown dwarfs are too neutral for magnetic fields to couple to the gas and produce magnetic spots \citep{2002ApJ...577..433G,2002ApJ...571..469M}}.   \citet{2000ASPC..212..322G} showed that Jupiter exhibits a 4.78 $\mu$m peak-to-trough intensity variation of 0.2 mag over its approximately 10 hr rotation period and the first detections of photometric variability in early-type L dwarfs were ascribed to the evolution of the cloud structures  \citep{2001A&A...367..218B,2002ApJ...577..433G}.

Variability in integrated light can also potentially be produced by thermal perturbations deep in the atmosphere \citep[e.g.,][]{2014ApJ...785..158R,2014ApJ...789L..14M}.   \citet{2014ApJ...788L...6Z} showed that depending on the relative strengths of the internal heat flux and radiative dissipation time scales, the surfaces of brown dwarfs will either be eddy-/turbulance or jet-dominated resulting in amplitude variations of a few percent.  Because the coolest brown dwarfs (i.e. Y dwarfs) have low internal heat fluxes and long dissipation time scales, \citeauthor{2014ApJ...788L...6Z} predict that Y dwarfs will exhibit zonal jets similar to those seen on Jupiter.

L and T dwarfs have been subjected to intense photometric and spectroscopic observations to search for variability at red optical (7000--10000 \AA), near-infrared (1--2.5 $\mu$m), and mid-infrared ($>$2.5 $\mu$m) wavelengths using both ground- and space-based observatories \citep[e.g.,][; see \citet{2017AstRv..13....1B} for a good review]{2012ApJ...760L..31B,2013ApJ...779..172G,2014ApJ...793...75R,2014A&A...566A.111W,2015ApJ...799..154M,2015MNRAS.448.3775R,2016ApJ...826....8Y,2017Sci...357..683A,2017ApJ...849..163S, 2019MNRAS.483..480V}.  Targeted ground-based near-infrared photometric surveys of L and T dwarfs show variability at the few percent level across the entire L and T sequence (variability fractions of 15 to 20\%) but found that high-amplitude variability ($>$2\%) occurs more frequently at the L/T transition.  The exact mechanism for the transition between the L and T dwarfs is currently unknown but one theory is that iron-, magnesium-, and silicon-bearing clouds break up at this point in the cooling sequence and so enhanced variability presumably due to the breakup of the clouds is consistent with these observations.  The largest variability survey of L and T dwarfs at mid-infrared wavelengths was a \textsl{Spitzer Space Telescope} Exploration Science program conducted by \citet{2015ApJ...799..154M}.  They used the Infrared Array Camera \citep[IRAC;][]{2004ApJS..154...10F} to observe 44 L3--T8 dwarfs at 3.6 and 4.5 $\mu$m (hereafter \cho\ and \cht).  They found variability with amplitudes ranging from the sub percent level to a few percent and computed a higher variability fractions of 61\%$^{+17\%}_{-20\%}$ and 31\%$^{+25\%}_{-17\%}$ for the L and T dwarfs, respectively.  

In contrast to the L and T dwarfs, the cooler Y dwarfs have not been studied in as much detail because they are difficult to observe with high precision at red optical and near-infrared wavelengths given how intrinsically faint they are these wavelengths \citep[$M_H >20$ mag;][]{2012ApJ...753..156K}.  \citet{2016ApJ...830..141L} observed WISEPA J173835.53$+$273259 in the $Y$ and $J$ bands for nearly 5 hr over two different nights but could not identify any statistically significant variability.  \citet{2015csss...18..997M} was the first to search for mid-infrared photometric variability in the Y dwarfs using AllWISE single-frame photometry \citep{2013wise.rept....1C} and found no obvious variability due to the large single-frame photometric uncertainty of $\sim$0.2 mag.  \citet{2016ApJ...823..152C} presented the first detection of photometric variability in the Y dwarf WISEPC J140518.40$+$553421.4 using \textit{Spitzer} at [3.6] and [4.5].  They observed the Y dwarf at two different epochs separated by 149 days and found variability at [4.5] with a semi-amplitude\footnote{The definition of ``amplitude'' varies in the literature.  In this and our previous work \citep{2016ApJ...823..152C,2016ApJ...830..141L}, we define the amplitude, $2A$, to be the difference between the peak and trough height (i.e. peak-to-trough) and the semi-amplitude, $A$, to be half the difference between the peak and trough height.} of 3.5\% and a period of 8.5 hr.  Moreover, they found that the light curves changed between epochs suggesting a change in the underlying cause or causes of the variability.  Variability at the same wavelengths and at similar semi-amplitudes has also been detected in multi-epoch observations of two other Y dwarfs WISEPA J173835.53$+$273259 \citep{2016ApJ...830..141L}, and WISE J085510.83$-$071442.5 \citep{2016ApJ...832...58E}.  The former shows periodic variability only at [4.5] while the latter shows periodic variability in both bands in the first epoch but more irregular variability in the second epoch.

The \textsl{Spitzer} observations of WISEPC J140518.40$+$553421.4 and WISEPA J173835.53$+$273259.0 were both part of a larger 550 hr \textsl{Spitzer Space Telescope} Exploration Science program to survey fourteen Y dwarfs for photometric variability and in this paper we present the results from the entire sample.  In \S2, we describe the observations and how we reduced the data while in \S3, we discuss how we determine which objects are variable and compute variability fractions.  Finally, in \S4 and 5, we discuss in detail the results of our variability analysis and then summarize the take away points of our Y dwarf survey.  

\section{Observations and Data Reduction}
\label{sec:obs}

\begin{deluxetable*}{lccccccc}
\tabletypesize{\scriptsize}
\tablecaption{Y Dwarf Targets\label{tbl:targets}}
\tablecolumns{8}
\tablewidth{0pt}
\tablehead{
\colhead{}& \colhead{} &
\colhead{}& \multicolumn{2}{c}{Epoch 1} &\colhead{}&
\multicolumn{2}{c}{Epoch 2}\\\cline{4-5}\cline{7-8}
\colhead{Object} &
\colhead{Spectral Type} & 
\colhead{Ref.} & \colhead{\textsl{Spitzer} AOR \#}&
\colhead{UT Date} &\colhead{}& \colhead{\textsl{Spitzer} AOR \#}& \colhead{UT Date} \\
\colhead{}& \colhead{} &
\colhead{}& \colhead{(\cho,\cht)} &\colhead{}&\colhead{}&
\colhead{(\cho,\cht)} &\colhead{}
}
\startdata
WISE J014656.66$+$423410.0AB  & T9/Y0     & 1 & (47160064,47172864) & 2013 Apr 19 && (47167744,47163392) & 2013 Oct 11\\ 
WISE J035000.32$-$565830.2    & Y1        & 2 & (47147264,47147776) & 2012 Dec 11 && (47172352,47168256) & 2013 Oct 08\\ 
WISE J035934.06$-$540154.6    & Y0        & 2 & (47170560,47165952) & 2013 Jan 04 && (47160832,47173632) & 2014 Apr 13\\ 
WISEPA J041022.71$+$150248.5  & Y0        & 3 & (46713088,46713856) & 2012 Nov 15 && (47169024,47164160) & 2013 Nov 03\\ 
WISE J053516.80$-$750024.9    & $\geq$Y1:  & 2 & (50483200,50482688) & 2014 May 25 && (50483968,50483456) & 2014 Sep 29\\ 
WISE J071322.55$-$291751.9    & Y0        & 2 & (47169280,47164672) & 2013 Jun 18 && (47158784,47171072) & 2014 Feb 07\\ 
WISE J073444.02$-$715744.0    & Y0        & 2 & (47158528,47172096) & 2013 Apr 28 && (47166464,47162368) & 2013 Aug 26\\ 
WD 0806$-$661B                & $\cdots$  & $\cdots$  & (49251328,49251072) & 2014 Aug 25 && ( $\cdots$,49252096) & 2014 Sep 09  \\
WISE J085510.83$-$071442.5    & Y4  & 4,5  & (52668160,52667904) & 2015 Mar 10/09 && (52668928,52668672) & 2015 Aug 02  \\
WISEPC J140518.40$+$553421.4  & Y0.5 pec? & 6,7 & (47166208,47162624) & 2013 Mar 22 && (47173888,47169792) & 2013 Aug 17\\ 
WISEPA J154151.66$-$225025.2  & Y1        & 6 & (47159808,47173120) & 2013 May 02 && (47168000,47163648) & 2013 Oct 06\\ 
WISEA J163940.84$-$684739.4   & Y0 pec    & 6 & (47165184,47161088) & 2013 Jun 09 && (47172608,47168768) & 2013 Oct 27\\ 
WISEPA J173835.53$+$273259.0  & Y0        & 3 & (47174144,47170048) & 2013 Jun 29 && (47164416,47159296) & 2013 Oct 29\\ 
WISEPA J182831.08$+$265037.8  & $\geq$Y2  & 8 & (46988032,46987776) & 2012 Nov 28 && (47157760,47170304) & 2013 Dec 14\\ 
WISEPC J205628.90$+$145953.3  & Y0        & 3 & (47171584,47166720) & 2013 Aug 19 && (47161600,47174400) & 2014 Jan 17\\ 
WISE J222055.32$-$362817.5    & Y0        & 2 & (47171840,47166976) & 2013 Aug 13 && (47162112,47174656) & 2014 Jan 30\\
\enddata
\tablecomments{References for spectral types: (1) \citet{2015ApJ...803..102D} (2) \citet{2012ApJ...753..156K} (3) \citet{2011ApJ...743...50C} (4) \citet{2019ApJS..240...19K} (5) \citet{2024AJ....167....5L}(6) \citet{2015ApJ...804...92S} (7) \citet{2016ApJ...823..152C} (8) \citet{2021ApJ...920...20C}. }
\end{deluxetable*}

Our \textsl{Spitzer} campaign (program \#90015) was designed to search for and characterize photometric variability of the fourteen spectroscopically confirmed Y dwarfs known at the time of the proposal's submission in 2012.  A log of the observations can be found in Table~\ref{tbl:targets}; hereafter we abbreviate WISE sources as WISE HHMM$+$DD.  One object, WISE 0146$+$42 was originally classified as Y0 by \citet{2012ApJ...753..156K} but was later found to be a T9/Y0 binary by \citet{2015ApJ...803..102D}.  We conducted our observations using IRAC at both [3.6] and [4.5]. Observing was conducted in ``staring mode'' whereby the target is kept close to the same position, in our case the so-called ``sweet spot'', on the array in order to minimize the effect that variations in quantum efficiency across an individual pixel has on the photometry \citep[i.e. the pixel phase effect;][]{2005PASP..117..978R}. Each Y dwarf was observed continuously for 24 hr--12 hr in \cho\  followed by 12 hr in \cht. The duration of 12 hr was motivated by four factors: $1)$ the rotation periods of L and T dwarfs measured from photometric varability were all below 10 hr \citep{2005ESASP.560..429B,2014A&amp;A...566A.130C}, $2)$ the upper limit to the rotation periods of T dwarfs from $v\sin(i)$ measurements is 12.5 hr \citep{2006ApJ...647.1405Z}, $3)$ the rotation period of Jupiter is $\sim$10 hr, and $4)$ the recommended maximum duration for staring mode by the \textsl{Spitzer} \textit{Science Center} was 12 hr.  The observations were repeated again between 122 and 464 days later in order to search for changes in the light curves on the timescale of months.

To this sample of 14 Y dwarfs we added WISE J085510.83$-$071442.5 \citep{2014ApJ...786L..18L} and WD 0806$-$661B \citep{2011ApJ...730L...9L, 2012ApJ...744..135L}, the only other Y dwarfs to be observed by \textsl{Spitzer} for photometric variability.  They were both observed in a similar manner to our program (i.e. multiple bands at two epochs) with two important differences:  1) observations at [4.5] were taken before [3.6] in each epoch, and 2) no second-epoch [3.6] observations were taken of WD 0806$-$661B.  The resulting light curves for WISE 0855$-$07 have already been analyzed by \citet{2016ApJ...832...58E} who found variability at both [3.6] and [4.5] at both epochs while the data for WD 0806$-$661B comes from program \#10031 (PI: Burgasser).

We reduced all of the data as described in \citet{2016ApJ...823..152C}, with a few minor differences in centroiding and background subtraction.  We began our data reduction by converting the Basic Calibrated Data (BCD) frames generated by the Spitzer Science Center using version S19.2.0 of the IRAC science pipeline from units of MJy sr$^{-1}$ to total electrons.  We did not apply the high-resolution gain map of the sweet spot pixel or the pixel phase correction because our targets sometimes fell up to two pixels from the sweet spot.  In order to identify the centroid position of our targets in each frame, we used the 2D Gaussian centroiding function from the \texttt{photutils} Python package.  As a initial guess for the routine, we used the centroid position determined from a median stack of the entire time series.  We then used the IDL \texttt{aper} routine with the \texttt{EXACT} keyword set to measure the total number of electrons within a 2 pixel radius of the centroid position.  In some cases there are background sources close to our target and so instead of using an annulus around our target to determine the background level, we identified a region on the frame devoid of sources to compute the median background level within a 7 pixel radius. We checked that the selection of the background region did not affect the intensity or shape of each light curve.  Clear outliers in the light curves, defined as data points where the total number of electrons exceed the median intensity level by more than 50 times the median absolute deviation (MAD=median$|$\{$\mathbf{x}$-median($\mathbf{x}$)\}$|$, were then removed before finally dividing by the median intensity level.  We note that one Y dwarf, WISE 1639$-$68, is located $\sim$3.3 (4$''$) and $\sim$4.5 (5$''$) pixels away from the $J$=14.90 star 2MASS 16394085$-$6847446 in epoch 1 and 2 respectively and as such, its light curve may be slightly contaminated by light from this star.  

%The normalized light curves of the Y0 dwarf WISE 0359$-$54 are shown in Figure~\ref{fig:W0359}.  The light curves not only exhibit clear variability at [4.5], but also a small fraction of bad data points.  

Figure \ref{fig:alllightcurves} shows the light curves of the 16 Y dwarfs at epochs 1 and 2, repsectively.  For display purposes only, we have identified and removed outliers as described in \S\ref{sec:bayes}.

%\begin{figure}
%\centering
%\includegraphics{WISEJ0359-5401.pdf}
%\caption{[3.6] (blue) and [4.5] (red) light curves for the Y0 dwarf WISE 0359$+$54 taken on two different dates.  The light curve observed on 2013 Jan 04 exhibits periodic variability at both wavelengths with a semi-amplitude of a few percent.  Note that the scale of the ordinate is different for [3.6] and [4.5] light curves.}
%\label{fig:W0359}
%\end{figure}

\begin{figure*}
\centerline{\hbox{\includegraphics{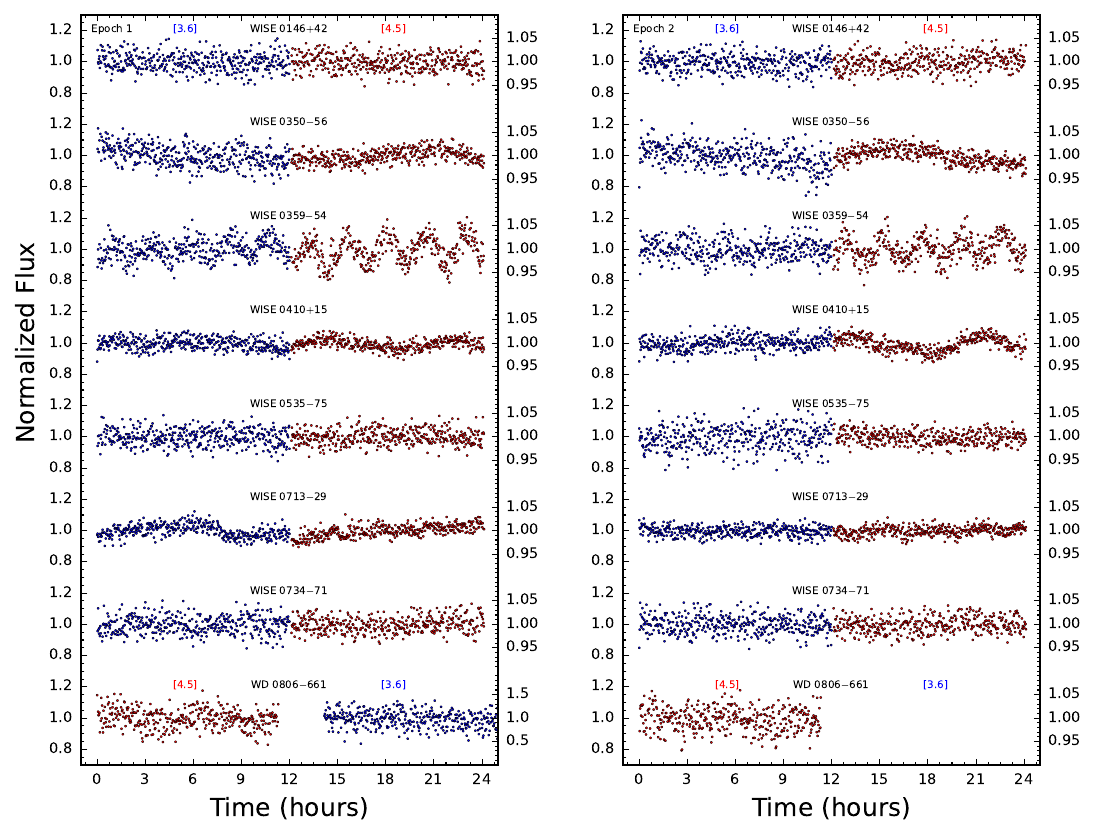}}}
\caption{Normalized IRAC \cho\ (blue) and \cht\ (red) photometry for the sixteen Y dwarfs in our sample. The epoch 1 data is found in the left panel and the epoch 2 data is found in the right panel.  For display purposes only, outliers have been removed as described in \S\ref{sec:bayes}. Note that the scale of the ordinate is different for [3.6] and [4.5] light curves.  \label{fig:alllightcurves}}
\end{figure*}

\setcounter{figure}{0}

\begin{figure*}
\centerline{\hbox{\includegraphics{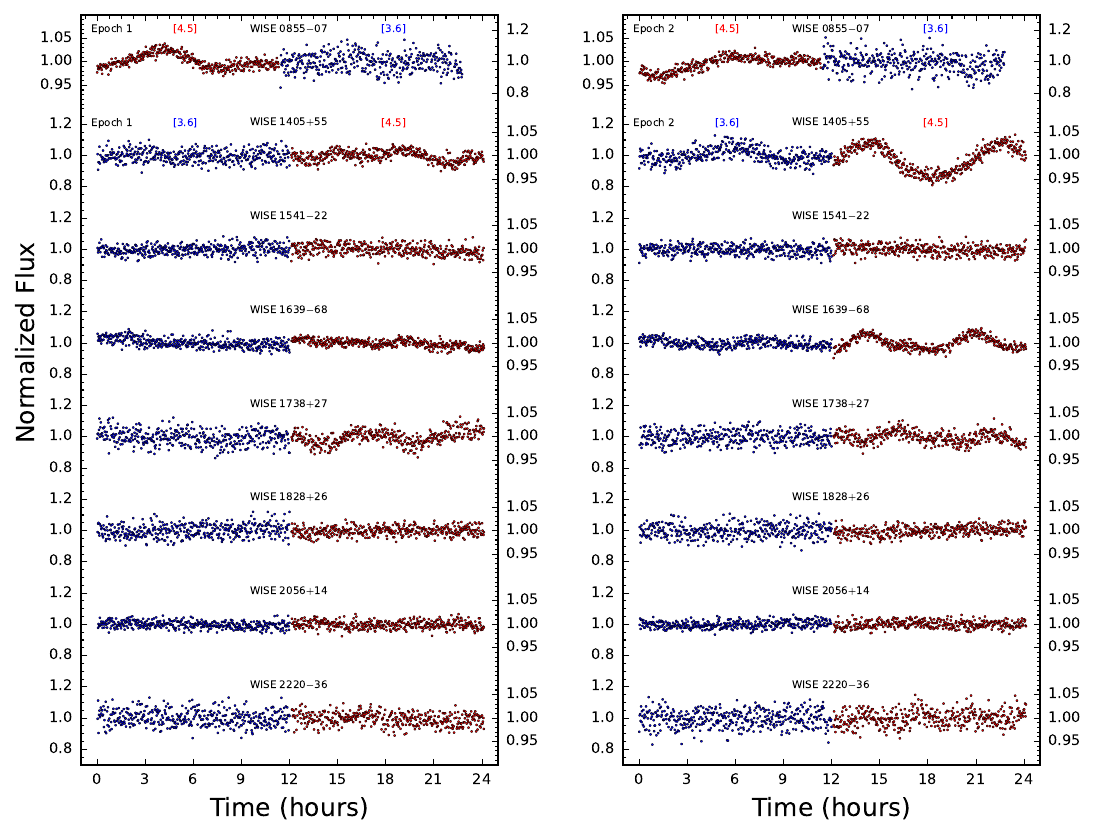}}}
  \caption{(Continued)}
\end{figure*}

\section{Variability Analysis}

\subsection{A Visual Inspection}

%\textcolor{orange}{The by-eye count:}\\
%\textcolor{orange}{[3.6] Epoch 1: 0350, 0359, 0713, 1405, 0855} \\
%\textcolor{orange}{[4.5] Epoch 1: 0350, 0359, 0410, 0713, 1405, 1738, 0855} \\
%\textcolor{orange}{[3.6] Epoch 2: 0350, 0359, 1405} \\
%\textcolor{orange}{[4.5] Epoch 2: 0350, 0359, 0410, 1405, 1639, 1738, 0855} \\

Inspection of the light curves shown in Figure \ref{fig:alllightcurves} reveal periodic and semi-periodic variability at the few percent level in many of the Y dwarf light curves.  By eye, we estimate a variability fraction of 31\% (5/16) and 44\% (7/16) at [3.6] and [4.5] in epoch 1 and 20\% (3/15) and 44\% (7/16) at [3.6] and [4.5] in epoch 2.  The [3.6] and [4.5] photometric bands sample similar pressures levels in the atmospheres of Y dwarfs \citep{2016ApJ...823..152C}, and so the most likely explanation for the differences in the by-eye variability fractions between bands is that we are not as sensitive to variability at [3.6] because Y dwarfs are $\sim$2.5 mag fainter in this band and the [3.6] and [4.5] exposure times are the same.  We defer the computation of a more quantitative estimate of the variability fractions until \S\ref{sec:varfrac}.  

One of the goals of our survey was to search for changes in Y dwarf light curves on timescales of months in order to investigate whether the light curves are stable or not.  While many of the light curves show subtle changes between the two epochs, three Y dwarfs exhibit clear changes to their light curves between epochs.  The first, WISE 0713$-$29, show a clear signature of variability at [3.6] in epoch 1 but the second-epoch [3.6] light curve obtained 234 days later shows no evidence of this variability.  A second Y dwarf, WISE 1405$-$55 which was discussed in detail in \citet{2016ApJ...823..152C}, exhibits both changes in the shape of the light curves and in the amplitudes of the variation.  The first-epoch observations only show low-level variability at [4.5] while the both the [3.6] and [4.5] light curves are variable in the second epoch.  In addition, the shape and amplitude of the [4.5] light curve has changed dramatically becoming more sinusoidal and having a larger amplitude.  The third Y dwarf, WISE 1639$-$68, shows the same pattern of changes between epochs that is exhibited by WISE 1405$-$55.  As noted in \S\ref{sec:obs}, WISE 1639$-$68 is located $\sim$4.5$''$ away from the $J$=14.90 star 2MASS 16394085$-$6847446 and as such, its light curve may be slightly contaminated by light from this star.  However, any contamination is unlikely to cause the changes seen between epochs.  Taken as a whole, \textsl{Spitzer} Y dwarf light curves are generally stable over timescales of months, but in some cases they can show clear variations in amplitude.

\subsection{Variability Fractions}
\label{sec:varfrac}

To provide a more rigorous estimate of the variability fraction in our sample, we use two different methods.  The first is described in \S\ref{sec:bayes} and is a Bayesian method inspired by the work of \citet{2012A&A...546A..89B} wherein we use the Bayesian Information Criterion \citep[hereafter BIC; ][]{schwarz1978} to identify which of three different variability models (a constant, a sine curve, and a double sine curve) fit the data the best.  The second is described in \S\ref{sec:freq} and uses the more traditional periodogram analysis to identify periodic signals in time series data \citep[e.g.,][]{2018ApJS..236...16V}.

\subsubsection{Bayesian Analysis}
\label{sec:bayes}

Our Bayesian analysis starts with the assumption that the data arise from the generative probabilistic model 
\begin{equation}
D_{i}=\mathcal{M}_{m,i} +\epsilon,
\end{equation}
where $D_{i}$ is a random variable for the number of electrons detected at time $t_{i}$, $\epsilon$ is random variable with a mean of zero and a variance of $\sigma^{2}$ that accounts 
for measurement uncertainty, and $\mathcal{M}_{m,i}$ is one of the following three models:
 
\begin{eqnarray}
\mathcal{M}_{1,i}&=&C\label{eq:cons},\\
%\mathcal{M}_{2,i}&=&mt_{i}+b\label{eq:line},\\
\mathcal{M}_{2,i}&=&A\sin\left(\frac{2\pi}{P}t_{i}+\phi\right)+C\label{eq:sine},\text{ and}\\
% \begin{aligned}
\mathcal{M}_{3,i}&=&A_{1}\sin\left(\frac{2\pi}{P}t_{i}+\phi_{1}\right) \\
& & +A_{2}\sin\left[2\left(\frac{2\pi}{P}\right)t_{i}+\phi_{2}\right] +C\label{eq:dsin},
% \end{aligned}
\end{eqnarray}

\noindent
where $C$ is a constant, and $A$, $P$, $\phi$ are the semi-amplitude, period, and phase of a sine curve.  We selected these three models because visually they seemed to encompass the variety of variability seen in our light curves.

If we denote the parameters of a given model as $\boldsymbol{\theta}$ (e.g., $\boldsymbol{\theta}$=[$A$, $P$, $\phi$] for model $\mathbfcal{M}_1$), then we can determine the joint probability distribution of the model parameters given our $N$ observations $\mathbf{d}=[d_{1},d_{2},d_{3},\cdots,d_{N}]$ using Bayes' theorem

\begin{equation}
p(\boldsymbol{\theta}|\mathbf{d})\propto\mathcal{L}(\mathbf{d}|\boldsymbol{\theta})p(\boldsymbol{\theta}),\label{eq:bayes}
\end{equation}

where $p(\boldsymbol{\theta}|\mathbf{d})$ is the posterior distribution, $p(\boldsymbol{\theta})$ is the prior function, and $\mathcal{L}(\mathbf{d} | \boldsymbol{\theta})$ is the likelihood.  We assume that the random variable $\epsilon$ follows a normal distribution with a mean of zero and a variance of $\sigma^2$, and so given that the data points are independent, we can write the likelihood function for the $m$th model as,

\begin{equation}
\label{eq:likelihood}
  \mathcal{L}_m(\boldsymbol d|\boldsymbol \theta, \sigma) = \left ( \frac{1}{\sqrt{2\pi \sigma^2}} \right )^N \text{exp} \left [ - \sum_{i=0}^N \left ( \frac{[d_i - \mathcal{M}_{m,i}(\boldsymbol \theta)]^2 }{2\sigma^2} \right ) \right ] .
\end{equation}

However as noted in \S\ref{sec:obs}, our data exhibit outliers and so we must account for these points in our likelihood function.  We therefore assume a combined generative model wherein the good data points are generated from the probability density function given by Equation \ref{eq:likelihood} while the outliers (or ``bad'') data points are generated from a normal distribution with a mean of $Y_{\text{bad}}$ and a variance of $\sigma_{\text{bad}}^{2}$ \citep{2010arXiv1008.4686H}.  The likelihood for our data set of $N$ observations is then given by,

\begin{equation}
\mathcal{L}_m\left(\mathbf{d}|\boldsymbol{\theta}\right)=\prod_{i=0}^{N}\left[\frac{1-P_{\text{bad}}}{\sqrt{2\pi\sigma^{2}}}\exp{\left(-\frac{\left[d_{i}-\mathcal{M}_{m,i}\right]^{2}}{2\sigma^{2}}\right)} \right. \nonumber
\end{equation}

\begin{equation}
\left. + \frac{P_{\text{bad}}}{\sqrt{2\pi\sigma_{\text{bad}}^{2}}}\exp{\left(-\frac{\left[d_{i}-Y_{\text{bad}}\right]^{2}}{2\sigma_{\text{bad}}^{2}}\right)}\right],\label{eq:like}
\end{equation}

%\begin{eqnarray}
%\mathcal{L}_m\left(\mathbf{d}|\boldsymbol{\theta}\right)&=&\prod_{i=0}^{N}\left[\frac{1-P_{\text{bad}}}{\sqrt{2\pi\sigma^{2}}}\exp{\left(-\frac{\left[d_{i}-\mathcal{M}_{m,i}\right]^{2}}{2\sigma^{2}}\right)} +\frac{P_{\text{bad}}}{\sqrt{2\pi\sigma_{\text{bad}}^{2}}}\exp{\left(-\frac{\left[d_{i}-Y_{\text{bad}}\right]^{2}}{2\sigma_{\text{bad}}^{2}}\right)}\right],\label{eq:like}
%\end{eqnarray}
\noindent
where $P_{\text{bad}}$ is the probability that a data point is bad (see \citet{2010arXiv1008.4686H} for derivation of this equation).

We sample the joint posterior distribution using Goodman \& Weare's Affine Invariant Markov Chain Monte Carlo (MCMC) Ensemble Sampler implemented by the Python program \texttt{emcee} \citep{2013PASP..125..306F}. We used 100 walkers to initialize the ensemble sampler along with 10500 steps, and discarded 50,000 samples from the initial burn-in leaving a total sample set of $1\times 10^6$.  We also used uniform (or ``uninformative'') priors for all parameters of the generative probabalistic models.  We then used the integrated autocorrelation time to determine if our MCMC chain has run long enough for each of the parameter chains to converge \citep{2018ApJS..236...11H}.  Following the blog post\footnote{\href{https://dfm.io/posts/autocorr/}{https://dfm.io/posts/autocorr/}} of Daniel Foreman-Mackey, we calculate the autocorrelation time at 10 points along our sampling chain.  If each parameter reaches $N>50\tau$, where $N$ is the number steps along the chain and $\tau$ is the autocorrelation time at that step then the model is considered converged.  As noted by \citet{2018ApJS..236...11H}, this is really only a heuristic test of convergence and so we also visually inspected the marginalized posterior distributions generated by the \textit{corner} routine and identified 5 additional models that fail to converge.  

We then identify which of the converged models best fit our data by assigning a score to each model and then comparing the scores to determine the best fit.  As our score we use the BIC which is defined as

\begin{equation}
\mathrm{BIC}=-2\mathcal{L}_{max} + k\log(n), \label{eq:bic}
\end{equation}

\noindent
where $\mathcal{L}_{max}$ is the maximized likelihood function, $k$ is the number of parameters in the model, and $n$ is the number of data points.  The latter term penalizes the BIC for overfitting by the increase in the number of model parameters.  We compare two models, $M_{1}$ and $M_{2}$ by calculating $\Delta$ BIC$_{12}$=BIC$_{1}-$BIC$_{2}$.  The $\Delta$BIC$_{12}$ values provide the evidence to favor $M_{2}$ over $M_{1}$: if $\Delta$ BIC$_{12}$ is $0-2$ then the difference is not worth mentioning, $2-6$ is positive evidence for $M_{2}$, $6-10$ is strong evidence for $M_2$, and any value greater than $10$ is very strong evidence for $M_2$ \citep{doi:10.1002/wics.199}. Table~\ref{tab:bic} gives the $\Delta$BIC values between the constant and sine models and the sine and double sine models.  The prior distributions used in the MCMC analysis and the 16, 50, and 84 percentile values of the parameters of the best fitting models for WISE 0146$+$4234 are given in Table \ref{tab:W0146}--\ref{tab:W2220}.  With the best fitting model in hand, we can now identify the outliers in the data following \citet{2010arXiv1008.4686H} whereby we compute the ratio of the second term in Equation \ref{eq:like} to the first term (the ratio of the ``bad'' part of the likelihood to the ``good'' part of the likelihood) and identify data points as outliers if this ratio exceeds 0.2.

We identify a source as variable if the $\Delta$BIC value between the constant and sine model and or the sine and double sine model is greater than 6 and the results are summarized in Table \ref{tab:varsum}.  We find a variability fraction of 19\% (3/16) and 56\% (9/16) at [3.6] and [4.5] in epoch 1 and 27\% (4/15) and 56\% (9/16) at [3.6] and [4.5] in epoch 2.  The model periods and the semi-amplitudes of the variable sources are given in Table \ref{tab:panda}.  The double sine models have semi-amplitudes for each sine curve and so we computed a single semi-amplitude for these models using a Monte Carlo technique.  We generated 10,000 models by drawing randomly from normal distributions for the model parameters $P$, $A_1$, $A_2$, $\phi_1$, $\phi_2$, and $C$.  The semi-amplitude is given by half the difference between the crest and trough heights and the values given in Table \ref{tab:panda} are the mean and standard deviation of the 10,000 values.  

\begin{deluxetable}{lRRRRRRRRRRRR}
\tablecaption{$\Delta$ BIC Values\tablenotemark{a} \label{tab:bic}}
\tablehead{
  \colhead{} & 
  \multicolumn{5}{c}{Epoch 1} & 
  \colhead{} & 
  \multicolumn{5}{c}{Epoch 2} \\
  \cline{2-6}  
  \cline{8-12}  
  \colhead{} &
  \multicolumn{2}{c}{[3.6]}& 
  \colhead{} & 
  \multicolumn{2}{c}{[4.5]} &
  \colhead{} & 
  \multicolumn{2}{c}{[3.6]} &
  \colhead{} & 
  \multicolumn{2}{c}{[4.5]} \\
  \cline{2-3}
  \cline{5-6}
  \cline{8-9}
  \cline{11-12}  
  \colhead{Object} &
  \colhead{c$-$s\tablenotemark{b}} & 
  \colhead{s$-$ds} & 
  \colhead{} &
  \colhead{c$-$s} & 
  \colhead{s$-$ds} & 
  \colhead{} &
  \colhead{c$-$s} & 
  \colhead{s$-$ds} & 
  \colhead{} &
  \colhead{c$-$s} & 
  \colhead{s$-$ds} 
}
\startdata
WISE J0146$+$42 & $\cdots$ & $\cdots$ & & $\cdots$ & $\cdots$ & & $\cdots$ & $\cdots$ & & $\cdots$ & $\cdots$\\ 
WISE J0350$-$56 & $\cdots$ & $\cdots$ & & $ 101.3$ & $\cdots$ & & $   6.1$ & $   8.9$ & & $ 219.8$ & $\cdots$\\ 
WISE J0359$-$54 & $  77.6$ & $\cdots$ & & $ 228.2$ & $\cdots$ & & $  14.3$ & $\cdots$ & & $ 166.1$ & $\cdots$\\ 
WISE J0410$+$15 & $\cdots$ & $\cdots$ & & $ 132.7$ & $\cdots$ & & $\cdots$ & $\cdots$ & & $ 224.9$ & $\cdots$\\ 
WISE J0535$-$75 & $\cdots$ & $\cdots$ & & $\cdots$ & $\cdots$ & & $\cdots$ & $\cdots$ & & $\cdots$ & $\cdots$\\ 
WISE J0713$-$29 & $ 136.4$ & $  18.5$ & & $  21.8$ & $\cdots$ & & $\cdots$ & $\cdots$ & & $  24.4$ & $   0.9$\\ 
WISE J0734$-$71 & $\cdots$ & $\cdots$ & & $\cdots$ & $\cdots$ & & $\cdots$ & $\cdots$ & & $  -0.2$ & $\cdots$\\ 
WD 0806$-$661B  & $\cdots$ & $\cdots$ &  \phn & $19.7$ & $\cdots$ &  \phn & $-$ & $-$ &  \phn & $\cdots$ &$\cdots$ \\
WISE J0855$+$07 & $1.0$    & $\cdots$ &  \phn & $379.4$ & $96.2$ &  \phn & $\cdots$ & $\cdots$ &  \phn & $  380.1$ & $\cdots$ \\
WISE J1405$+$55 & $\cdots$ & $\cdots$ & & $  76.4$ & $  56.8$ & & $ 146.4$ & $\cdots$ & & $ 715.0$ & $  34.1$\\ 
WISE J1541$-$22 & $\cdots$ & $\cdots$ & & $\cdots$ & $\cdots$ & & $\cdots$ & $\cdots$ & & $\cdots$ & $\cdots$\\ 
WISE J1639$-$68 & $\cdots$ & $\cdots$ & & $  50.9$ & $  51.9$ & & $  69.4$ & $\cdots$ & & $ 333.2$ & $ 112.6$\\ 
WISE J1738$+$27 & $   9.3$ & $\cdots$ & & $ 169.5$ & $   0.4$ & & $\cdots$ & $\cdots$ & & $  59.9$ & $  28.3$\\ 
WISE J1828$+$26 & $\cdots$ & $\cdots$ & & $\cdots$ & $\cdots$ & & $\cdots$ & $\cdots$ & & $\cdots$ & $\cdots$\\ 
WISE J2056$+$14 & $\cdots$ & $\cdots$ & & $\cdots$ & $\cdots$ & & $\cdots$ & $\cdots$ & & $\cdots$ & $\cdots$\\ 
WISE J2220$-$36 & $\cdots$ & $\cdots$ & & $\cdots$ & $\cdots$ & & $\cdots$ & $\cdots$ & & $  18.4$ & $\cdots$\\ 
\enddata
\tablenotetext{a}{$\cdots$ indicates the sine and double-sine curves did not converge (see \S\ref{sec:bayes}).  $-$ indicates no data at those wavelengths.}
\tablenotetext{b}{c=constant, s=sine, ds=double}
\end{deluxetable}

\begin{deluxetable}{lRRRRRRRRRRRR}
\tablecaption{Periods and Semi-Amplitudes for Variable-Source Models from Bayesian Analysis\label{tab:panda}}
\tablehead{
  \colhead{} & 
  \multicolumn{5}{c}{Epoch 1} & 
  \colhead{} & 
  \multicolumn{5}{c}{Epoch 2} \\
  \cline{2-6}  
  \cline{8-12}  
  \colhead{} &
  \multicolumn{2}{c}{[3.6]}& 
  \colhead{} & 
  \multicolumn{2}{c}{[4.5]} &
  \colhead{} & 
  \multicolumn{2}{c}{[3.6]} &
  \colhead{} & 
  \multicolumn{2}{c}{[4.5]} \\
  \cline{2-3}
  \cline{5-6}
  \cline{8-9}
  \cline{11-12}  
  \colhead{Object} &
  \colhead{$P$ (hr)} & 
  \colhead{$A$ (\%)} & 
  \colhead{} &
  \colhead{$P$ (hr)} & 
  \colhead{$A$ (\%)} & 
  \colhead{} &
  \colhead{$P$ (hr)} & 
  \colhead{$A$ (\%)} & 
  \colhead{} &
  \colhead{$P$ (hr)} & 
  \colhead{$A$ (\%)} 
}
\startdata
WISEJ0350$-$56 & $\cdots$             & $\cdots$    & & $11.4^{+0.5}_{-0.4}$    & $1.07\pm0.09$ & & $6.2^{+0.2}_{-0.1}$   & 4.0$\pm$0.4\tablenotemark{a}          & & $14.0^{+0.9}_{-0.7}$  & $1.55\pm0.09$ \\ 
WISEJ0359$-$54 & $2.41^{+0.02}_{-0.03}$ & 4.2$\pm$0.4 & & 2.45$\pm$0.01       & $2.8\pm0.1$   & & $2.44^{+0.06}_{-0.05}$ & 2.5$\pm$0.5 & & $2.44\pm0.02$        & $2.2\pm0.2$ \\ 
WISEJ0410$+$15 & $\cdots$             & $\cdots$    & & $7.8\pm0.2$         & $0.92\pm0.07$ & & $\cdots$             & $\cdots$    & & $8.2\pm0.2$          & $1.59^{+0.09}_{-0.08}$ \\ 
WISEJ0713$-$29 & $10.1\pm0.3$         & 4.0$\pm$0.2\tablenotemark{a} & & $6.5\pm0.2$         & $0.58\pm0.09$ & & $\cdots$             & $\cdots$    & & $3.15^{+0.07}_{-0.06}$ & $0.44\pm0.07$ \\ 
WD 0806$-$661B & $\cdots$ & $\cdots$ &  \phn & $5.7^{+0.3}_{-0.2}$ & $2.7\pm0.5$ &  \phn & $-$ & $-$ &  \phn & $\cdots$ & $\cdots$ \\
WISE J0855$+$07 &  $\cdots$ &  $\cdots$ &  \phn & $11.5^{+0.5}_{-0.7}$ & $1.94\pm0.8$\tablenotemark{a} &  \phn &  $\cdots$ &  $\cdots$ &  \phn & $  17\pm 1$ & $2.0^{+0.2}_{-0.1}$\\
WISEJ1405$+$55 & $\cdots$             & $\cdots$    & & $9.1^{+0.4}_{-0.3}$   & 1.08$\pm$0.06\tablenotemark{a}           & & $8.4^{+0.3}_{-0.2}$   & $3.7\pm0.3$ & & $8.42^{+0.06}_{-0.05}$  & 3.52$\pm$0.04\tablenotemark{a} \\ 
WISEJ1639$-$68 & $\cdots$             & $\cdots$    & & $6.32^{+0.09}_{-0.08}$ & 0.68$\pm$0.04\tablenotemark{a}          & & $6.8\pm0.2$          & $1.6\pm0.2$ & & $6.69\pm0.05$        & 1.75$\pm$0.06\tablenotemark{a} \\ 
WISEJ1738$+$27 & $4.8\pm0.2$          & $1.7\pm0.3$ & & $5.91^{+0.10}_{-0.09}$ & $1.3\pm0.09$ & & $\cdots$             & $\cdots$    & & $6.09\pm0.09$        & 1.08$\pm$0.07\tablenotemark{a} \\ 
WISEJ2220$-$36 & $\cdots$             & $\cdots$    & & $\cdots$             & $\cdots$     & & $\cdots$             & $\cdots$    & & $3.23\pm0.08$        & $0.7\pm0.1$ \\ 
\enddata
\tablenotetext{a}{The light curve is fit best by a double sine curve.  The semi-amplitude is derived as described in \S\ref{sec:bayes}.}
\end{deluxetable}

\subsubsection{Periodogram Analysis}\label{sec:freq}

\begin{figure*}
\centerline{\hbox{\includegraphics{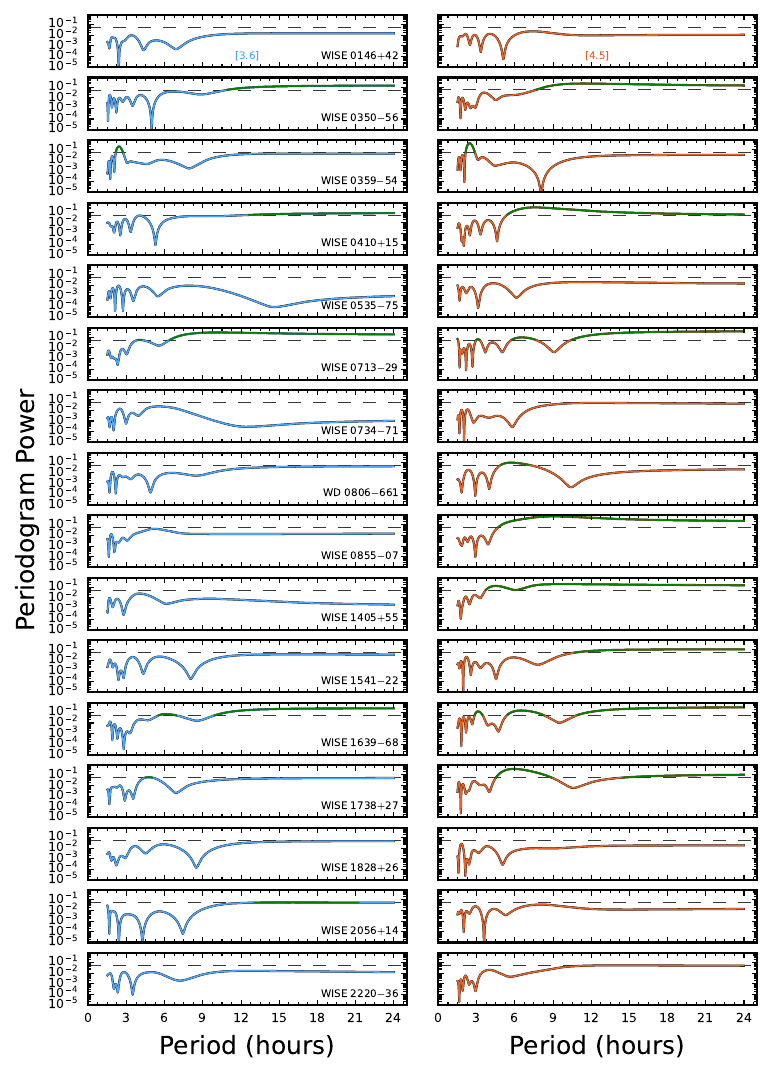}}}
  \caption{Epoch 1 periodograms for the 16 Y dwarfs in our sample. The grey dashed lines give the 5\% (95\% confidence) False Alarm Levels which give the power levels above which we would expect to see power less than 5\% of the time under the null hypothesis of pure Gaussian noise (no variable signal).   Any object with power in its periodogram above this level is considered variable.\label{fig:periodograms1}}
\end{figure*}

\begin{figure*}
\centerline{\hbox{\includegraphics{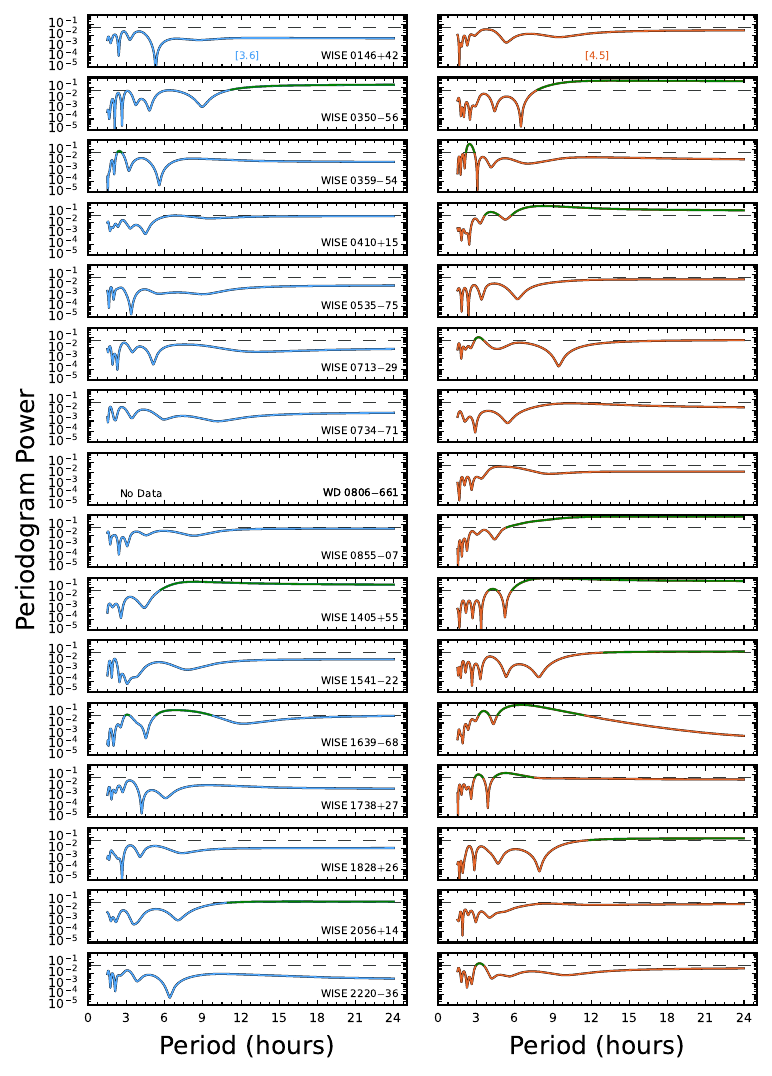}}}
  \caption{Epoch 2 periodograms for the 16 Y dwarfs in our sample. The grey dashed lines give the 5\% (95\% confidence) False Alarm Levels which give the power levels above which we would expect to see power less than 5\% of the time under the null hypothesis of pure Gaussian noise (no variable signal).   Any object with power in its periodogram above this level is considered variable.\label{fig:periodograms2}}

\end{figure*}

The second method we used to identify variable sources is to construct periodograms of the light curves using the \texttt{LombScargle} routine in the Python \texttt{astropy.stats} library.  We first removed data points wherein the normalized intensity exceeded unity by more than $\pm5\times$ the median absolute deviation of the light curve to ensure those outliers did not skew the results.  We then set our frequency grid by taking the inverse of an evenly space array of periods ranging from 1.5 to 24 hr with a step of 0.1 hr.  Figures \ref{fig:periodograms1} and \ref{fig:periodograms2} show the epoch 1 and 2 periodograms for the 16 Y dwarfs, respectively.  Also shown are the 5\% (95\% confidence) False Alarm Levels (computed using the \citet{2008MNRAS.385.1279B} approximation) which give the power levels above which we would expect to see power less than 5\% of the time under the null hypothesis of pure Gaussian noise (no variable signal).   Any object with power in its periodogram above this level is considered variable and a summary of which Y dwarfs are variable in each band and epoch is given in Table \ref{tab:varsum}.  We find a variability fraction of 44\% (7/16) and 63\% (8/16) at [3.6] and [4.5] in epoch 1 and 33\% (5/15) and 63\% (10/16) at [3.6] and [4.5] in epoch 2.

\begin{deluxetable}{lccccccccccc}
\tablecaption{Y Dwarf Variability Summary\label{tab:varsum}}
\tablehead{
%  \colhead{} &
%%  \multicolumn{5}{c}{By Eye} &   
  \colhead{} &
  \multicolumn{5}{c}{Bayesian Information Criterion} & 
  \colhead{} &
  \multicolumn{5}{c}{False Alarm Probability}\\
  \cline{2-6}
  \cline{8-12}
%  \cline{14-18}  
  \colhead{} & 
  \multicolumn{2}{c}{Epoch 1} &
  \colhead{} & 
  \multicolumn{2}{c}{Epoch 2} & 
  \colhead{} & 
  \multicolumn{2}{c}{Epoch 1} &
  \colhead{} & 
  \multicolumn{2}{c}{Epoch 2} \\
  \cline{2-3}
  \cline{5-6}
  \cline{8-9}
  \cline{11-12}
  \colhead{Object} & 
  \colhead{[3.6]} & 
  \colhead{[4.5]} & 
  \colhead{} & 
  \colhead{[3.6]} & 
  \colhead{[4.5]} &  
  \colhead{} & 
  \colhead{[3.6]} & 
  \colhead{[4.5]} & 
  \colhead{} & 
  \colhead{[3.6]} & 
  \colhead{[4.5]}   
}
\startdata
WISE J0146$+$42AB & N & N &&  N & N && N & N && N & N \\ 
WISE J0350$-$56   & N & Y &&  Y & Y && Y & Y && Y & Y \\ 
WISE J0359$-$54   & Y & Y &&  Y & Y && Y & Y && Y & Y \\ 
WISE J0410$+$15   & N & Y &&  N & Y && Y & Y && N & Y \\ 
WISE J0535$-$75   & N & N &&  N & N && N & N && N & N \\ 
WISE J0713$-$29   & Y & Y &&  N & Y && Y & Y && N & N \\ 
WISE J0734$-$71   & N & N &&  N & N && N & N && N & N \\ 
WD 0806$-$661B    & N & Y &&  $-$ & N && N & Y && $-$ & N \\
WISE J0855$-$07   & N & Y &&  N & Y && N & Y && N & Y \\
WISE J1405$+$55   & N & Y &&  Y & Y && N & Y && Y & Y \\ 
WISE J1541$-$22   & N & N &&  N & N && N & Y && N & Y \\ 
WISE J1639$-$68   & N & Y &&  Y & Y && Y & Y && Y & Y \\ 
WISE J1738$+$27   & Y & Y &&  N & Y && Y & Y && N & Y \\ 
WISE J1828$+$26   & N & N &&  N & N && N & N && N & Y \\ 
WISE J2056$+$14   & N & N &&  N & N && Y & N && Y & N \\ 
WISE J2220$-$36   & N & N &&  N & Y && N & N && N & Y \\ 
\hline              
Fraction          &3/16 (19\%)  &9/16 (56\%) & & 4/15 (27\%)  & 9/16 (56\%) & & 7/16 (44\%) & 8/16 (63\%) & & 5/15 (33\%) & 10/16 (63\%) \\ 
\enddata
\end{deluxetable}

\subsection{Variability Fraction Confidence Intervals}
\label{sec:varstat}

To provide confidence intervals for the variability fractions we derived in the previous two sections, we follow the Bayesian formalism originally devised by \citet{2007ApJ...670.1367L} to compute exoplanet occurrence rates but which has recently been used by \citet{2019MNRAS.483..480V} to compute brown dwarf variability fractions.  If we define $f$ to be the fraction of objects in our survey that exhibit variability with semi-amplitudes and periods in the interval [0.5\%, 10\%]$\cap$[1.5 hr, 20 hr] in a given epoch and in a given band, then we can use Bayes' theorem to derive the posterior probability distribution for $f$ given the data $\{d_{i}\}_{i=1}^{16}$, where $d_i=1$ means variability was detected in the light curve of the $i$th object and $d_i=0$ means variability was not detected in the light curve of the $i$th object (see Table \ref{tab:varsum} wherein Y=1 and N=0).

The probability of detecting variability in the $i$th object is $f p_i$, where $p_i$ is the probability that such variability would be detected in the semi-amplitude and period interval [0.5\%, 10\%]$\cap$[1.5 hr, 20 hr] given the observations.  It then follows that the probability of not detecting variability in the $i$th target is $(1-f p_i)$.  Since the probability of detecting variability in a given target is independent of the other targets, the likelihood function for the entire dataset $\{d_{i}\}_{i=1}^{16}$ can be written as,

\begin{equation}
\label{eq:varlike}
\mathcal{L}\left(\{d_{i}\}_{i=1}^{16}| f\right)=\prod_{i=1}^{16} \left(1-f p_{i}\right)^{1-d_{i}} \left(f p_{i}\right)^{d_{i}}.
\end{equation}

\noindent
The posterior distribution for $f$ is then given by Bayes theorem,

\begin{equation}
\label{eq:varprob}
p\left(f|\{d_{i}\}_{i=1}^{16}\right)=\frac{\mathcal{L}\left(\{d_{i}\}_{i=1}^{16}| f\right)p\left(f\right)}{\int_{0}^{1}\mathcal{L}\left(\{d_{i}\}_{i=1}^{16}| f\right)p\left(f\right)\text{d}f}.
\end{equation}

\begin{figure*}[!h]
\centering
\includegraphics{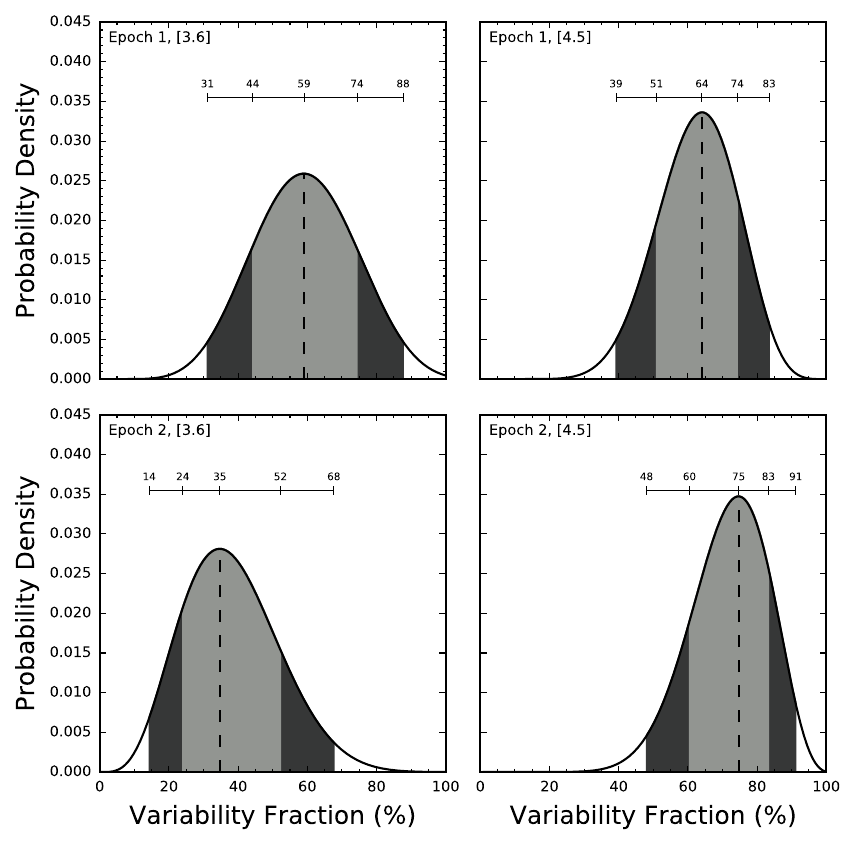}
\caption{Posterior distributions for the variability fraction $f$ for each epoch and each band.  68\% and 95\% central credibility intervals are indicated in light grey and dark grey, respectively. }
\label{fig:vpost}
\end{figure*}

In order to evaluate the posterior distribution, we require knowledge of $p_i$ for each target in each band and at each epoch and the prior distribution $p(f)$. As noted above, $p_i$ is the probability that variability would be detected in the $i$th target given the observations.  We compute this value for each light curve in a Monte Carlo fashion wherein we generate 10,000 variable light curves with random phase shifts, semi-amplitudes, and periods and then determine $p_i$ by computing the fraction of these light curves identified as variable.  For purely computational efficiency, we choose to use periodograms to identify which sources are variable.  We generated sine curves with phase shifts, semi-amplitudes, and periods drawn from random distributions given by $\mathcal{U}(0,2\pi$), $\mathcal{U}$(0.5,10), and $\mathcal{U}$(1.5, 20).  Noise for each curve was generated by drawing randomly from a normal distribution given by $\mathcal{N}(0,\sigma^2$), where $\sigma$ is the standard deviation of the best fitting model for that light curve.  The $p_i$ values for each target in each band and epoch is given in Table \ref{tab:pi}.  With the $p_i$ values in hand and assuming a uniform prior for $p(f)$, we evaluate the posterior distribution for $f$ in each band and epoch using Equations \ref{eq:varlike} and \ref{eq:varprob} and the results are shown in Figure \ref{fig:vpost}; 68\% and 95\% central credibility intervals are indicated in grey.

\begin{deluxetable}{lccccc}
\tablecaption{Survey Completeness\label{tab:pi}}
\tablehead{
  \colhead{} & 
  \multicolumn{2}{c}{Epoch 1} &
  \colhead{} & 
  \multicolumn{2}{c}{Epoch 2} \\
  \cline{2-3}
  \cline{5-6}
  \colhead{Object} & 
  \colhead{[3.6]} & 
  \colhead{[4.5]} & 
  \colhead{} & 
  \colhead{[3.6]} & 
  \colhead{[4.5]}}
\startdata
WISE J0146$+$42AB & 0.76 &  0.94 &&  0.76 &  0.95   \\
WISE J0350$-$56   & 0.70 &  0.96 &&  0.68 &  0.97   \\
WISE J0359$-$54   & 0.72 &  0.91 &&  0.71 &  0.92   \\
WISE J0410$+$15   & 0.88 &  0.99 &&  0.84 &  0.96   \\
WISE J0535$-$75   & 0.73 &  0.97 &&  0.67 &  0.96   \\
WISE J0713$-$29   & 0.85 &  0.98 &&  0.89 &  1.00   \\
WISE J0734$-$71   & 0.77 &  0.89 &&  0.69 &  0.95   \\
WD 0806$-$661B    & 0.06 &  0.62 &&  $-$  &  0.50 \\
WISE J0855$-$07   & 0.65 &  0.99 &&  0.63 &  0.99   \\
WISE J1405$+$55   & 0.84 &  0.99 &&  0.84 &  0.92   \\
WISE J1541$-$22   & 0.89 &  0.99 &&  0.86 &  0.99   \\
WISE J1639$-$68   & 0.90 &  1.00 &&  0.92 &  0.98   \\
WISE J1738$+$27   & 0.81 &  0.98 &&  0.80 &  0.98   \\
WISE J1828$+$26   & 0.85 &  1.00 &&  0.81 &  1.00   \\
WISE J2056$+$14   & 0.93 &  1.00 &&  0.92 &  1.00   \\
WISE J2220$-$36   & 0.78 &  0.98 &&  0.77 &  0.97   \\
\enddata
\end{deluxetable}

\section{Discussion}

\subsection{Variability Fractions}

The Bayesian and periodogram analyses find that 1) a significant fraction of the Y dwarfs are variable and 2) the observed Y dwarf variability fraction is larger at [4.5] than at [3.6].  However, the periodogram analysis identified 10 light curves as variable that the Bayesian analysis does not.  This is primarily because the periodograms often show increasing power to longer periods indicating long-term variability which the Bayesian analysis does not capture.  The discrepant variability fractions underscores the fact that whether variability is detected or not may depend heavily on the technique used and as a result, comparisons between variability fractions derived from different surveys can be challenging.  The inferred variability fractions shown in Figure \ref{fig:vpost} all indicate high variability fractions ranging from 35\% up to 75\%.  The [3.6] fractions disagree at the 1$\sigma$ level while the [4.5] fractions are formally consistent.

\begin{figure}[!h]
\centering
\includegraphics[width=3.5in]{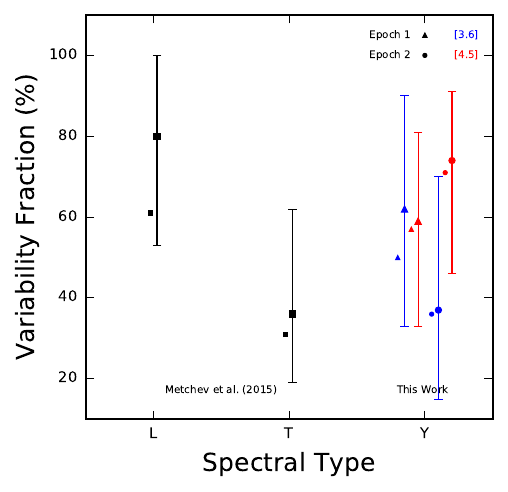}
\caption{\textit{Spitzer} variability fractions for L and T dwarfs \citep{2015ApJ...799..154M} and the Y dwarfs (this work).  The points without error bars are the raw fractions (number of variable objects/total sample) while the points with error bars have been corrected for survey sensitivity limits.  The error bars for the L and T dwarfs are 95\% confidence limits while the error bars for the Y dwarfs are centered 95\% credibility intervals.  Note that the detection thresholds of this work and that of \citeauthor{2015ApJ...799..154M} are different.}  
\label{fig:varcomp}
\end{figure}

With variability fraction credibility intervals in hand, we can compare our results to previous surveys in order to place our Y dwarf survey into context.  The survey that is most similar to ours is the aforementioned \citeauthor{2015ApJ...799..154M} \textit{Spitzer} Exploration Science program which surveyed 44 L3--T8 dwarfs at [3.6] and [4.5].  Their survey included a 14 hr observation of each dwarf at [3.6] followed by a 7 hr observation at [4.5] and so while not identical to our observing strategy, it is similar enough that our results can be compared.  Figure \ref{fig:varcomp} shows the variability fractions for the two different surveys.  While the values are consistent within the uncertainties, the point estimates for the fractions show a clear drop for the T dwarfs.  While it is generally assumed that variations in the horizontal and/or vertical cloud structures gives rise to the observed variability, it has proven difficult to prove definitively \citep[e.g.,][]{2016ApJ...832...58E}.  In broad terms, the drop in the variability fraction in the T dwarfs actually supports the idea that clouds are the primary cause of the observed variability.  In the standard paradigm of brown dwarf evolution, L dwarfs become progressively redder at near-infrared wavelengths due to the presence of silicate and liquid iron condensate clouds.  The early to mid-type T dwarfs are blue because these clouds are disrupted by some as-yet unknown mechanism \citep{2002ApJ...571L.151B,2004AJ....127.3553K}.  In the late-type T dwarfs and Y dwarfs, additional condensate clouds composed of KCl, and Na$_2$S and even water ice form \citep{2000ASPC..212..152M,2012ApJ...756..172M}.  The cloudy$\rightarrow$cloud free$\rightarrow$cloudy nature of the spectral sequence is weakly supported by the high$\rightarrow$low$\rightarrow \sim$high variability fraction in the L, T, and Y dwarfs.

\subsection{Rotation Periods}

Five Y dwarfs show clear periodic variation in at least 2 light curves and have model periods that are consistent within 2$\sigma$.  We therefore assume that the rotation periods of these Y dwarfs are given by the model periods.  The weighted means of these periods (computed using the larger of the two uncertainties associated with a given period), which range from 2.44 to 8.42 hr, are given in Table \ref{tab:rotper}.  Three additional Y dwarfs, WISE 0350$-$56, WD 0806$-$661B, and WISE 2220$-$36 show periodic variations but we cannot be sure they correspond to their rotation periods.  In the case of WISE 0350$-$56, the model derived periods of 11.4$^{+0.5}_{-0.4}$ hr and 14.0$^{+0.9}_{-0.7}$ hr are clearly inconsistent while periods for WD 0806$-$661B and WISE 2220$-$36 could only be computed in a single band.  We designate the periods of these three Y dwarfs in Table \ref{tab:rotper} as tentative.  Three of the light curves for a third Y dwarf, WISE 0713$-$29, are fit by sine curves with disparate periods (10.1, 6.5, and 3.15 hr) and so we do not include it in Table \ref{tab:rotper}.  The disparate periods underscore the fact even though a model may describe a given data set well, it may not have any physical meaning.

\begin{deluxetable}{lcl}[t]
\tablecaption{Rotation Periods of Y Dwarfs\label{tab:rotper}}
\tablehead{
  \colhead{Object} & 
  \colhead{Rotation Period (hr)} &
  \colhead{Input Epochs and Bands\tablenotemark{a}}}
\startdata
\multicolumn{3}{c}{Secure Periods} \\
\hline
WISE 0359$-$54   & 2.44$\pm$0.01 &E1 [3.6], E1 [4.5], E2 [3.6], E2 [4.5] \\  
WISE 0410$+$15   & 8.0$\pm$0.14 & E1 [4.5], E2 [4.5] \\ 
WISE 1405$+$55   & 8.42$\pm$0.06 & E2 [3.6], E2 [4.5] \\ 
WISE 1639$-$68   & 6.69$\pm$0.05 & E2 [3.6], E2 [4.5] \\  
WISE 1738$+$27   & 6.01$\pm$0.07 & E1 [4.5], E2 [4.5] \\ 
\hline
\multicolumn{3}{c}{Tentative Periods} \\
\hline
WISE 0350$-$56   & 11.7$\pm$0.44 & E1 [4.5], E2 [4.5] \\ 
WD 0806$-$661B   & 5.7$^{+0.3}_{-0.2}$ & E1 [4.5] \\
WISE 2220$-$36   & 3.23$\pm$0.8 & E2 [4.5] \\
\enddata
\tablenotetext{a}{E1=epoch 1, E2=epoch 2}
\end{deluxetable}

\subsection{Semi-Amplitudes}

\begin{figure}
\centering
\includegraphics{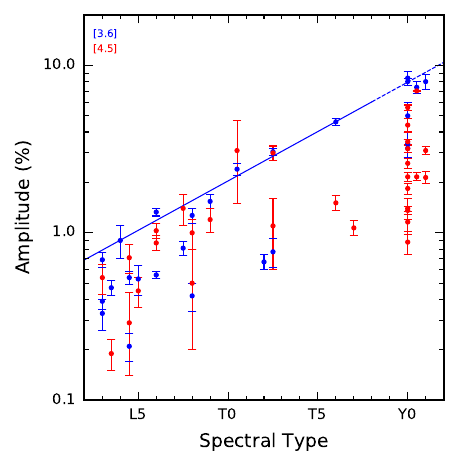}
\caption{Reproduction of Figure 9 from \citet{2015ApJ...799..154M} with the amplitudes of the Y dwarfs (twice the semi-amplitudes reported in Table \ref{tab:panda}) from this work included.  We have not included the upper limits for L and T dwarfs in the original figure.}
\label{fig:ampvspt}
\end{figure}

\citet{2015ApJ...799..154M} noted that the maximum observed amplitude at [3.6] and [4.5] increases with decreasing spectral type.  Beyond a spectral type of T3, however, the trend is driven by a single T6 dwarf, 2MASS J22282889$-$4310262.  With our sample of Y dwarfs, we can test whether this trend continues to later spectral types and cooler effective temperatures as 2MASS 2228$-$43 and Jupiter's 20\% variability at 4.78 $\mu$m suggests.   Figure \ref{fig:ampvspt} shows the amplitude as a function of spectral type for the L and T dwarfs in the \citeauthor{2015ApJ...799..154M} sample and the Y dwarfs (twice the semi-amplitudes reported in Table \ref{tab:panda}) in our sample along with the line (in the log-linear space) that \citeauthor{2015ApJ...799..154M} found to represent the envelope of maximum variability.  The Y dwarfs all fall on or below the extension of the \citeauthor{2015ApJ...799..154M} line with WISE 0350$-$56, WISE 0359$-$54, WISE 0713$-$29, and WISE 1405$+$55, falling nearly on it.  Our observations therefore confirm the trend continues to at least a spectral type of early Y.  In addition, large-amplitude variations ($>3$\%) are much more common amoung the Y dwarfs than the L and T dwarfs.   Finally, amoungst the Y dwarfs, the 4 largest amplitudes are from [3.6] light curves even though there are more [4.5] variable light curves.  The exact physical mechanism or mechanisms responsible for the trend however, remain unknown at this time.

%\textcolor{red}{Figure \ref{fig:ampratio} shows the ratio of the [4.5] to [3.6] semi-amplitudes as a function of spectral type for L, T, and Y dwarfs from \citet{2015ApJ...799..154M} and this work.  \citeauthor{2015ApJ...799..154M} found the mean amplitude to be 1 with a standard deviation of 0.7.  The weighted mean of this ratio derived from Y dwarf light curves with consistent periods is 0.81$\pm$0.04 which is formally inconsistent with the LT dwarf value.  Two additional Y dwarfs are plotted, WISE 0350$-$56 and WISE 1738$+$27, but the periods of the fits are formally inconsistent and so we do not include them in our mean value.  \citet{2016ApJ...823..152C} noted that the near unity ratio for WISE 1405$-$55 ruled out hot spots as a possible cause of the observed variability because the \citet{2014ApJ...789L..14M} hot spots models have ratios less than unity.  While a detailed study is beyond the scope of this work, the larger sample of Y dwarfs suggest that hot spots could account for the observed variability.  }

%\begin{figure}
%\centering
%\includegraphics{amp_ratio.pdf}
%\caption{Reproduction of Figure 10 from \citet{2015ApJ...799..154M} with the Y dwarfs from this work included.  \textcolor{red}{We only included those Y dwarfs for which we believed...}}
%\label{fig:ampratio}
%\end{figure}

\subsection{Phases}

\citet{2012ApJ...760L..31B} found that the 1.1--1.7 $\mu$m spectroscopic observations of L and T dwarfs with the \textit{Hubble Space Telescope} and [4.5] photometric observations with \textit{Spitzer} showed variability with a common period but with phases that were a function of wavelength \citep[see also][]{2016ApJ...826....8Y}.  Since different wavelengths probe different layers of an atmosphere, they found that the phase lags increased with decreasing pressure level (i.e. higher in the atmosphere).  Their most plausible interpretation involves a heterogeneous deep cloud controlling the near-infrared variability with horizontal temperature variations at higher pressure levels driving the longer wavelength variability.  No such joint observations have been conducted to date on Y dwarfs but we do have one Y dwarf in our sample whose light curves are fit with sine curves in both bands and both epochs, with which we can explore whether phase variations are present between the [3.6] and [4.5] bands.   WISE 0359$-$54 has the shortest period in our sample at 2.44 hr and shows strong variability in all four light curves.  The phases of the best fit sine curves are all consistent within their uncertainties suggesting no phase difference exists between [3.6] and [4.5].  This is perhaps not surprising as \citet{2016ApJ...823..152C} has shown that the [3.6] and [4.5] bands probe very similar, low-pressure layers ($P$=0.7--10 bar) of a Y dwarfs atmosphere.  WISE 0359$-$59 is therefore a perfect candidate for variability studies with the \textit{James Webb Space Telescope} because its NIRSpec instrument covers the 0.7--5 $\mu$m wavelength range in a single exposure.

\section{Summary}

We have conducted a search for varaibility in the mid-infrared light curves of Y dwarfs using the \textsl{Spitzer Space Telescope}.

\begin{enumerate}
\item We find that variability is common amongst Y dwarfs with variability fractions ranging from 35\% to 75\%.

\item While mid-infrared Y dwarf light curves are generally stable on time scales of months, three Y dwarfs showed clear, by-eye changes to their light curves between epoch 1 and epoch 2. 

\item The variability fractions of \textit{Spitzer} variability studies of L, T, and Y dwarfs (\citet{2015ApJ...799..154M} and this work) (weakly) support the standard paradigm that clouds are responsible for the observed variability.

\item We have determined the rotation period of five Y dwarfs with three additional tentative periods and they range from 2.44 hr to 8.42 hr.

\item We confirm that the trend in maximum amplitude as a function of spectral type found by \citet{2015ApJ...799..154M} extends at least to the early-type Y dwarfs.

%\item We find an average [4.5] to [3.6] amplitude ratio less than unity which suggests that hot spots may be the physical mechanism for the observed variability.

\item We find no phase difference between the [3.6] and [4.5] light curves in WISE 0359$-$54 which is perhaps unsurprising given the two bands sample similar atmospheric levels.  

\end{enumerate}

Finally, we note that these observations can be used as baselines for both photometric and spectroscopic studies of Y dwarf variability carried out by the James Webb Space Telescope (GO 11740, GO 2327).

\begin{acknowledgments}

This publication makes use of data products from the Wide-field Infrared Survey Explorer, which is a joint project of the University of California, Los Angeles, and the Jet Propulsion Laboratory/California Institute ofTechnology, funded by the National Aeronautics and Space Administrations and is based [in part] on observations made with the Spitzer Space Telescope, which is operated by the Jet Propulsion Laboratory, California Institute of Technology under a contract with NASA. Support for this work was provided by NASA through an award issued by JPL/Caltech.  This research has made use of the NASA/IPAC Infrared Science Archive, which is operated by the Jet Propulsion Laboratory, California Institute of Technology, under contract with the National Aeronautics and Space Administration.

\end{acknowledgments}

%\begin{contribution}
%%This section gives authors the space to recognize author contributions. The text inside this environment is NOT counted towards the total word quanta. At a minimum, manuscripts are expected to include this text:

%All authors contributed equally to the Terra Mater collaboration.

%% But authors are expected to provide more specific details, e.g. 
%%
%%SC was responsible for writing and submitting the manuscript.
%%WWM came up with the initial research concept and edited the manuscript.
%%OTS obtained the funding and edited the manuscript.
%%EBF provided the formal analysis and validation. He also edited the manuscript.
%%GEH Supervised the undergraduates, wrote the software and administers the project github and Zenodo repositories.
%%
%% Authors can use the Contributor Role Taxonomy (CRediT) at
%% for ideas on how write a good statement tailored to their needs.

%\end{contribution}

\facilities{Spitzer(IRAC)}

%\software{astropy \citep{2013A&A...558A..33A,2018AJ....156..123A,2022ApJ...935..167A},  
%          Cloudy \citep{2013RMxAA..49..137F}, 
%          Source Extractor \citep{1996A&AS..117..393B}
%          }


\pagebreak
\newpage

\bibliography{/Users/michaelcushing/Science/PaperDatabase/ref}
\bibliographystyle{./aasjournalv7}

\appendix

\begin{deluxetable}{llc}[!ht]
\tablecaption{WISEJ0146$+$4234 \label{tab:W0146}}
\tablecolumns{3}
\tablewidth{1.0\columnwidth}
\tablehead{
\colhead{Model Parameter} & \colhead{Prior\tablenotemark{a}} & \colhead{Value\tablenotemark{b}}
}
\startdata
\multicolumn{3}{c}{Epoch 1 \cho\ constant model parameters}\\ \hline
Constant $C$ & $\mathcal{U}(C>0)$  & 0.999$\pm$0.003 \\ 
Standard deviation $\sigma$ & $\mathcal{U}(0,1)$  & 0.056$\pm$0.002 \\ 
Bad data probability $P_{bad}$ & $\mathcal{U}(0,1)$  & 0.05$\pm$0.01 \\ 
Bad data standard deviation $\sigma_{bad}$ & $\mathcal{U}(0,4)$  & $0.56^{+0.11}_{-0.08}$ \\ 
Bad data mean $Y_{bad}$ & $\mathcal{U}(0.5,3.5)$  & $1.52^{+0.14}_{-0.13}$ \\ \hline 
\multicolumn{3}{c}{Epoch 1 \cht\ constant model parameters}\\ \hline
Constant $C$ & $\mathcal{U}(C>0)$  & 0.9989$\pm$0.0009 \\ 
Standard deviation $\sigma$ & $\mathcal{U}(0,1)$  & $0.0183^{+0.0008}_{-0.0007}$ \\ 
Bad data probability $P_{bad}$ & $\mathcal{U}(0,1)$  & 0.6$\pm$0.01 \\ 
Bad data standard deviation $\sigma_{bad}$ & $\mathcal{U}(0,4)$  & $0.17^{+0.03}_{-0.02}$ \\ 
Bad data mean $Y_{bad}$ & $\mathcal{U}(0.5,3.5)$  & 1.18$\pm$0.04 \\ \hline 
\multicolumn{3}{c}{Epoch 2 \cho\ constant model parameters}\\ \hline
Constant $C$ & $\mathcal{U}(C>0)$  & 0.994$\pm$0.003 \\ 
Standard deviation $\sigma$ & $\mathcal{U}(0,1)$  & 0.055$\pm$0.002 \\ 
Bad data probability $P_{bad}$ & $\mathcal{U}(0,1)$  & 0.05$\pm$0.01 \\ 
Bad data standard deviation $\sigma_{bad}$ & $\mathcal{U}(0,4)$  & $0.54^{+0.10}_{-0.08}$ \\ 
Bad data mean $Y_{bad}$ & $\mathcal{U}(0.5,3.5)$  & 1.4$\pm$0.1 \\ \hline 
\multicolumn{3}{c}{Epoch 2 \cht\ constant model parameters}\\ \hline
Constant $C$ & $\mathcal{U}(C>0)$  & 0.9993$\pm$0.0009 \\ 
Standard deviation $\sigma$ & $\mathcal{U}(0,1)$  & 0.0182$\pm$0.0007 \\ 
Bad data probability $P_{bad}$ & $\mathcal{U}(0,1)$  & 0.06$\pm$0.01 \\ 
Bad data standard deviation $\sigma_{bad}$ & $\mathcal{U}(0,4)$  & $0.20^{+0.04}_{-0.03}$ \\ 
Bad data mean $Y_{bad}$ & $\mathcal{U}(0.5,3.5)$  & 1.12$\pm$0.04 \\ \hline 
\enddata
\tablenotetext{a}{$\mathcal{U}(a,b)$ denotes a uniform distribution over the range $a$ to $b$.}
\tablenotetext{b}{The values reported correspond to the 16th, 50th, and 84th percentiles of the marginalized posterior distribution.}
\end{deluxetable}
\newpage

%%%%%%%%%%%%%%%%%%%%%%%%%%%%%%%%%%%%%%%%%%%% W0350-5658 %%%%%%%%%%%%%%%%%%%%%%%%%%%%%%%%%%%%%%%%%%%%

\begin{deluxetable}{llc}[!ht]
\tiny
\tablecaption{WISEJ0350$-$56 \label{tab:W0350}}
\tablecolumns{3}
\tablewidth{1.0\columnwidth}
\tablehead{
\colhead{Model Parameter} & \colhead{Prior\tablenotemark{a}} & \colhead{Value\tablenotemark{b}}
}
\startdata
\multicolumn{3}{c}{Epoch 1 \cho\ constant model parameters}\\ \hline
Constant $C$ & $\mathcal{U}(C>0)$  & 1.000$\pm$0.003 \\ 
Standard deviation $\sigma$ & $\mathcal{U}(0,1)$  & 0.066$\pm$0.003 \\ 
Bad data probability $P_{bad}$ & $\mathcal{U}(0,1)$  & 0.05$\pm$0.01 \\ 
Bad data standard deviation $\sigma_{bad}$ & $\mathcal{U}(0,4)$  & $0.60^{+0.12}_{-0.09}$ \\ 
Bad data mean $Y_{bad}$ & $\mathcal{U}(0.5,3.5)$  & $1.4^{+0.2}_{-0.1}$ \\ \hline 
\multicolumn{3}{c}{Epoch 1 \cht\ sine model parameters}\\ \hline
Semi-Amplitude $A$ ($\%$) & $\mathcal{U}(0,0.5)$  & 1.07$\pm$0.09 \\ 
Period $P$ (hr) & $\mathcal{U}(0,20)$  & $11.4^{+0.5}_{-0.4}$ \\ 
Phase $\phi$ (degrees) & $\mathcal{U}(-90,450)$  & $ 165^{+25}_{-24}$ \\ 
Constant $C$ & $\mathcal{U}(0.8,1.2)$  & 0.9992$\pm$0.0006 \\ 
Standard deviation $\sigma$ & $\mathcal{U}(0,1)$  & 0.0122$\pm$0.0005 \\ 
Bad data probability $P_{bad}$ & $\mathcal{U}(0,1)$  & 0.07$\pm$0.01 \\ 
Bad data standard deviation $\sigma_{bad}$ & $\mathcal{U}(0,4)$  & 0.13$\pm$0.02 \\ 
Bad data mean $Y_{bad}$ & $\mathcal{U}(0.5,3.5)$  & 1.14$\pm$0.03 \\ \hline 
\multicolumn{3}{c}{Epoch 2 \cho\ double sine model parameters}\\ \hline
Semi-Amplitude 1 $A_{1}$ ($\%$) & $\mathcal{U}(0,0.5)$  & 2.5$\pm$0.5 \\ 
Semi-Amplitude 2 $A_{2}$ ($\%$) & $\mathcal{U}(0,0.5)$  & 2.2$\pm$0.5 \\ 
Period $P$ (hr) & $\mathcal{U}(0,20)$  & $6.2^{+0.2}_{-0.1}$ \\ 
Phase 1 $\phi_{1}$ (degrees) & $\mathcal{U}(-90,450)$  & $ 7^{+16}_{-15}$ \\ 
Phase 2 $\phi_{2}$ (degrees) & $\mathcal{U}(-90,450)$  & $37^{+23}_{-22}$ \\ 
Constant $C$ & $\mathcal{U}(0.8,1.2)$  & 0.994$\pm$0.004 \\ 
Standard deviation $\sigma$ & $\mathcal{U}(0,1)$  & 0.071$\pm$0.003 \\ 
Bad data probability $P_{bad}$ & $\mathcal{U}(0,1)$  & 0.02$\pm$0.01 \\ 
Bad data standard deviation $\sigma_{bad}$ & $\mathcal{U}(0,4)$  & $0.9^{+0.3}_{-0.2}$ \\ 
Bad data mean $Y_{bad}$ & $\mathcal{U}(0.5,3.5)$  & 2.1$\pm$0.3 \\ \hline
\multicolumn{3}{c}{Epoch 2 \cht\ sine model parameters}\\ \hline
Semi-Amplitude $A$ ($\%$) & $\mathcal{U}(0,0.5)$  & 1.55$\pm$0.09 \\ 
Period $P$ (hr) & $\mathcal{U}(0,20)$  & $14.0^{+0.9}_{-0.7}$ \\ 
Phase $\phi$ (degrees) & $\mathcal{U}(-90,450)$  & $ 39^{+25}_{-22}$ \\ 
Constant $C$ & $\mathcal{U}(0.8,1.2)$  & $0.9975^{+0.0008}_{-0.0009}$ \\ 
Standard deviation $\sigma$ & $\mathcal{U}(0,1)$  & 0.0121$\pm$0.0005 \\ 
Bad data probability $P_{bad}$ & $\mathcal{U}(0,1)$  & 0.07$\pm$0.01 \\ 
Bad data standard deviation $\sigma_{bad}$ & $\mathcal{U}(0,4)$  & $0.16^{+0.03}_{-0.02}$ \\ 
Bad data mean $Y_{bad}$ & $\mathcal{U}(0.5,3.5)$  & 1.13$\pm$0.03 \\ 
\enddata
\tablenotetext{a}{$\mathcal{U}(a,b)$ denotes a uniform distribution over the range $a$ to $b$.}
\tablenotetext{b}{The values reported correspond to the 16th, 50th, and 84th percentiles of the marginalized posterior distribution.}
\end{deluxetable}
\newpage
%%%%%%%%%%%%%%%%%%%%%%%%%%%%%%%%%%%%%%%%%%%% W0359-5401 %%%%%%%%%%%%%%%%%%%%%%%%%%%%%%%%%%%%%%%%%%%%

\begin{deluxetable}{llc}[!ht]
\tiny
\tablecaption{WISEJ0359$-$54 \label{tab:W0359}}
\tablecolumns{3}
\tablewidth{1.0\columnwidth}
\tablehead{
\colhead{Model Parameter} & \colhead{Prior\tablenotemark{a}} & \colhead{Value\tablenotemark{b}}
}
\startdata
\multicolumn{3}{c}{Epoch 1 \cho\ sine model parameters}\\ \hline
Semi-Amplitude $A$ ($\%$) & $\mathcal{U}(0,0.5)$  & 4.2$\pm$0.4 \\ 
Period $P$ (hr) & $\mathcal{U}(0,20)$  & $2.41^{+0.02}_{-0.03}$ \\ 
Phase $\phi$ (degrees) & $\mathcal{U}(-90,450)$  & 293$\pm$ 11 \\ 
Constant $C$ & $\mathcal{U}(0.8,1.2)$  & 0.999$\pm$0.003 \\ 
Standard deviation $\sigma$ & $\mathcal{U}(0,1)$  & 0.056$\pm$0.002 \\ 
Bad data probability $P_{bad}$ & $\mathcal{U}(0,1)$  & 0.05$\pm$0.01 \\ 
Bad data standard deviation $\sigma_{bad}$ & $\mathcal{U}(0,4)$  & 0.7$\pm$0.1 \\ 
Bad data mean $Y_{bad}$ & $\mathcal{U}(0.5,3.5)$  & 1.8$\pm$0.2 \\ \hline 
\multicolumn{3}{c}{Epoch 1 \cht\ sine model parameters}\\ \hline
Semi-Amplitude $A$ ($\%$) & $\mathcal{U}(0,0.5)$  & 2.8$\pm$0.1 \\ 
Period $P$ (hr) & $\mathcal{U}(0,20)$  & 2.45$\pm$0.01 \\ 
Phase $\phi$ (degrees) & $\mathcal{U}(-90,450)$  & $315^{+16}_{-17}$ \\ 
Constant $C$ & $\mathcal{U}(0.8,1.2)$  & 0.9973$\pm$0.0008 \\ 
Standard deviation $\sigma$ & $\mathcal{U}(0,1)$  & 0.0197$\pm$0.0008 \\ 
Bad data probability $P_{bad}$ & $\mathcal{U}(0,1)$  & 0.10$\pm$0.02 \\ 
Bad data standard deviation $\sigma_{bad}$ & $\mathcal{U}(0,4)$  & $0.29^{+0.04}_{-0.03}$ \\ 
Bad data mean $Y_{bad}$ & $\mathcal{U}(0.5,3.5)$  & 1.15$\pm$0.05  \\ \hline 
\multicolumn{3}{c}{Epoch 2 \cho\ sine model parameters}\\ \hline
Semi-Amplitude $A$ ($\%$) & $\mathcal{U}(0,0.5)$  & 2.5$\pm$0.5 \\ 
Period $P$ (hr) & $\mathcal{U}(0,20)$  & $2.44^{+0.06}_{-0.05}$ \\ 
Phase $\phi$ (degrees) & $\mathcal{U}(-90,450)$  & $356^{+24}_{-23}$ \\ 
Constant $C$ & $\mathcal{U}(0.8,1.2)$  & 1.000$\pm$0.003 \\ 
Standard deviation $\sigma$ & $\mathcal{U}(0,1)$  & $0.061^{+0.003}_{-0.002}$ \\ 
Bad data probability $P_{bad}$ & $\mathcal{U}(0,1)$  & 0.05$\pm$0.01 \\ 
Bad data standard deviation $\sigma_{bad}$ & $\mathcal{U}(0,4)$  & 0.6$\pm$0.1 \\ 
Bad data mean $Y_{bad}$ & $\mathcal{U}(0.5,3.5)$  & $1.4^{+0.2}_{-0.1}$ \\ \hline  
\multicolumn{3}{c}{Epoch 2 \cht\ sine model parameters}\\ \hline
Semi-Amplitude $A$ ($\%$) & $\mathcal{U}(0,0.5)$  & 2.2$\pm$0.2 \\ 
Period $P$ (hr) & $\mathcal{U}(0,20)$  & 2.44$\pm$0.02 \\ 
Phase $\phi$ (degrees) & $\mathcal{U}(-90,450)$  & $347^{+20}_{-19}$ \\ 
Constant $C$ & $\mathcal{U}(0.8,1.2)$  & 1.001$\pm$0.001 \\ 
Standard deviation $\sigma$ & $\mathcal{U}(0,1)$  & 0.0202$\pm$0.0008 \\ 
Bad data probability $P_{bad}$ & $\mathcal{U}(0,1)$  & 0.05$\pm$0.01 \\ 
Bad data standard deviation $\sigma_{bad}$ & $\mathcal{U}(0,4)$  & $0.27^{+0.05}_{-0.04}$ \\ 
Bad data mean $Y_{bad}$ & $\mathcal{U}(0.5,3.5)$  & 1.18$\pm$0.06 \\ 
\enddata
\tablenotetext{a}{$\mathcal{U}(a,b)$ denotes a uniform distribution over the range $a$ to $b$.}
\tablenotetext{b}{The values reported correspond to the 16th, 50th, and 84th percentiles of the marginalized posterior distribution.}
\end{deluxetable}
\newpage
%%%%%%%%%%%%%%%%%%%%%%%%%%%%%%%%%%%%%%%%%%%% W0410+1502 %%%%%%%%%%%%%%%%%%%%%%%%%%%%%%%%%%%%%%%%%%%%

\begin{deluxetable}{llc}[!ht]
\tablecaption{WISEJ0410$+$15 \label{tab:W0410}}
\tablecolumns{3}
\tablewidth{1.0\columnwidth}
\tablehead{
\colhead{Model Parameter} & \colhead{Prior\tablenotemark{a}} & \colhead{Value\tablenotemark{b}}
}
\startdata
\multicolumn{3}{c}{Epoch 1 \cho\ constant model parameters}\\ \hline
Constant $C$ & $\mathcal{U}(C>0)$  & 0.996$\pm$0.002 \\ 
Standard deviation $\sigma$ & $\mathcal{U}(0,1)$  & 0.037$\pm$0.001 \\ 
Bad data probability $P_{bad}$ & $\mathcal{U}(0,1)$  & 0.04$\pm$0.01 \\ 
Bad data standard deviation $\sigma_{bad}$ & $\mathcal{U}(0,4)$  & $0.36^{+0.09}_{-0.06}$ \\ 
Bad data mean $Y_{bad}$ & $\mathcal{U}(0.5,3.5)$  & 1.5$\pm$0.1 \\ \hline 
\multicolumn{3}{c}{Epoch 1 \cht\ sine model parameters}\\ \hline
Semi-Amplitude $A$ ($\%$) & $\mathcal{U}(0,0.5)$  & 0.92$\pm$0.07 \\ 
Period $P$ (hr) & $\mathcal{U}(0,20)$  & 7.8$\pm$0.2 \\ 
Phase $\phi$ (degrees) & $\mathcal{U}(-90,450)$  & 135$\pm$0.2 \\ 
Constant $C$ & $\mathcal{U}(0.8,1.2)$  & 0.9968$\pm$0.0005 \\ 
Standard deviation $\sigma$ & $\mathcal{U}(0,1)$  & $0.0089^{+0.0004}_{-0.0003}$ \\ 
Bad data probability  $P_{bad}$ & $\mathcal{U}(0,1)$  & 0.07$\pm$0.01 \\ 
Bad data standard deviation $\sigma_{bad}$ & $\mathcal{U}(0,4)$  & 0.13$\pm$0.02 \\ 
Bad data mean $Y_{bad}$ & $\mathcal{U}(0.5,3.5)$  & 1.10$\pm$0.03 \\ \hline 
\multicolumn{3}{c}{Epoch 2 \cho\ constant model parameters}\\ \hline
Constant $C$ & $\mathcal{U}(C>0)$  & 0.999$\pm$0.002 \\ 
Standard deviation $\sigma$ & $\mathcal{U}(0,1)$  & 0.041$\pm$0.002 \\ 
Bad data probability $P_{bad}$ & $\mathcal{U}(0,1)$  & 0.05$\pm$0.01 \\ 
Bad data standard deviation $\sigma_{bad}$ & $\mathcal{U}(0,4)$  & $0.45^{+0.09}_{-0.07}$ \\ 
Bad data mean $Y_{bad}$ & $\mathcal{U}(0.5,3.5)$  & 1.5$\pm$0.1 \\ \hline 
\multicolumn{3}{c}{Epoch 2 \cht\ sine model parameters}\\ \hline
Semi-Amplitude $A$ ($\%$) & $\mathcal{U}(0,0.5)$  & $1.59^{+0.09}_{-0.08}$ \\ 
Period $P$ (hr) & $\mathcal{U}(0,20)$  & 8.2$\pm$0.2 \\ 
Phase $\phi$ (degrees) & $\mathcal{U}(-90,450)$  & 222$\pm$17 \\ 
Constant $C$ & $\mathcal{U}(0.8,1.2)$  & 0.9948$\pm$0.0006 \\ 
Standard deviation $\sigma$ & $\mathcal{U}(0,1)$  & $0.0111^{+0.0005}_{-0.0004}$ \\ 
Bad data probability $P_{bad}$ & $\mathcal{U}(0,1)$  & $0.07^{+0.02}_{-0.01}$ \\ 
Bad data standard deviation $\sigma_{bad}$ & $\mathcal{U}(0,4)$  & 0.14$\pm$0.02 \\ 
Bad data mean $Y_{bad}$ & $\mathcal{U}(0.5,3.5)$  & 1.10$\pm$0.03 \\ 
\enddata
\tablenotetext{a}{$\mathcal{U}(a,b)$ denotes a uniform distribution over the range $a$ to $b$.}
\tablenotetext{b}{The values reported correspond to the 16th, 50th, and 84th percentiles of the marginalized posterior distribution.}
\end{deluxetable}
\newpage
%%%%%%%%%%%%%%%%%%%%%%%%%%%%%%%%%%%%%%%%%%%% W0535-7500 %%%%%%%%%%%%%%%%%%%%%%%%%%%%%%%%%%%%%%%%%%%%

\begin{deluxetable}{llc}[!ht]
\tablecaption{WISEJ0535$-$75 \label{tab:W0535}}
\tablecolumns{3}
\tablewidth{1.0\columnwidth}
\tablehead{
\colhead{Model Parameter} & \colhead{Prior\tablenotemark{a}} & \colhead{Value\tablenotemark{b}}
}
\startdata
\multicolumn{3}{c}{Epoch 1 \cho\ constant model parameters}\\ \hline
Constant $C$ & $\mathcal{U}(C>0)$  & 0.999$\pm$0.003 \\ 
Standard deviation $\sigma$ & $\mathcal{U}(0,1)$  & 0.054$\pm$0.002 \\ 
Bad data probability $P_{bad}$ & $\mathcal{U}(0,1)$  & 0.05$\pm$0.01 \\ 
Bad data standard deviation $\sigma_{bad}$ & $\mathcal{U}(0,4)$  & 0.6$\pm$0.1 \\ 
Bad data mean $Y_{bad}$ & $\mathcal{U}(0.5,3.5)$  & 1.5$\pm$0.2 \\ \hline 
\multicolumn{3}{c}{Epoch 1 \cht\ constant model parameters}\\ \hline
Constant $C$ & $\mathcal{U}(C>0)$  & 1.0002 $\pm$0.0008 \\ 
Standard deviation $\sigma$ & $\mathcal{U}(0,1)$  & 0.0161$\pm$0.0006 \\ 
Bad data probability $P_{bad}$ & $\mathcal{U}(0,1)$  & 0.05$\pm$0.01 \\ 
Bad data standard deviation $\sigma_{bad}$ & $\mathcal{U}(0,4)$  & $0.17^{+0.04}_{-0.03}$ \\ 
Bad data mean $Y_{bad}$ & $\mathcal{U}(0.5,3.5)$  & 1.13$\pm$0.04 \\ \hline 
\multicolumn{3}{c}{Epoch 2 \cho\ constant model parameters}\\ \hline
Constant $C$ & $\mathcal{U}(C>0)$  & 0.991$\pm$0.004 \\ 
Standard deviation $\sigma$ & $\mathcal{U}(0,1)$  & 0.076$\pm$0.003 \\ 
Bad data probability $P_{bad}$ & $\mathcal{U}(0,1)$  & 0.06$\pm$0.01 \\ 
Bad data standard deviation $\sigma_{bad}$ & $\mathcal{U}(0,4)$  & 0.8$\pm$0.1 \\ 
Bad data mean $Y_{bad}$ & $\mathcal{U}(0.5,3.5)$  & 1.8$\pm$0.2 \\ \hline 
\multicolumn{3}{c}{Epoch 2 \cht\ constant model parameters}\\ \hline
Constant $C$ & $\mathcal{U}(C>0)$  & 0.9993$\pm$0.0007 \\ 
Standard deviation $\sigma$ & $\mathcal{U}(0,1)$  & 0.0141$\pm$0.0006 \\ 
Bad data probability $P_{bad}$ & $\mathcal{U}(0,1)$  & $0.08^{+0.02}_{-0.01}$ \\ 
Bad data standard deviation $\sigma_{bad}$ & $\mathcal{U}(0,4)$  & $0.19^{+0.03}_{-0.02}$ \\ 
Bad data mean $Y_{bad}$ & $\mathcal{U}(0.5,3.5)$  & $1.10^{+0.04}_{-0.03}$ \\ \hline 
\enddata
\tablenotetext{a}{$\mathcal{U}(a,b)$ denotes a uniform distribution over the range $a$ to $b$.}
\tablenotetext{b}{The values reported correspond to the 16th, 50th, and 84th percentiles of the marginalized posterior distribution.}
\end{deluxetable}
\newpage
%%%%%%%%%%%%%%%%%%%%%%%%%%%%%%%%%%%%%%%%%%%% W0713-2917 %%%%%%%%%%%%%%%%%%%%%%%%%%%%%%%%%%%%%%%%%%%%

\begin{deluxetable}{llc}[!ht]
\tablecaption{WISEJ0713$-$29 \label{tab:W0713}}
\tablecolumns{3}
\tablewidth{1.0\columnwidth}
\tablehead{
\colhead{Model Parameter} & \colhead{Prior\tablenotemark{a}} & \colhead{Value\tablenotemark{b}}
}
\startdata
\multicolumn{3}{c}{Epoch 1 \cho\ double sine model parameters}\\ \hline
Semi-Amplitude 1 $A_{1}$ ($\%$) & $\mathcal{U}(0,.5)$  & 3.4$\pm$0.2 \\ 
Semi-Amplitude 2 $A_{2}$ ($\%$) & $\mathcal{U}(0,.5)$  & 1.3$\pm$0.2 \\ 
Period $P$ (hr) & $\mathcal{U}(0,20)$  & 10.1$\pm$0.3 \\ 
Phase 1 $\phi_{1}$ (degrees) & $\mathcal{U}(-90,450)$  & 275$\pm$8 \\ 
Phase 2 $\phi_{2}$ (degrees) & $\mathcal{U}(-90,450)$  & $349^{+19}_{-17}$ \\ 
Constant $C$ & $\mathcal{U}(0.8,1.2)$  & 0.999$\pm$0.002 \\ 
Standard deviation $\sigma$ & $\mathcal{U}(0,1)$  & 0.031$\pm$0.001 \\ 
Bad data probability $P_{bad}$ & $\mathcal{U}(0,1)$  & 0.06$\pm$0.01 \\ 
Bad data standard deviation $\sigma_{bad}$ & $\mathcal{U}(0,4)$  & $0.42^{+0.07}_{-0.06}$ \\ 
Bad data mean $Y_{bad}$ & $\mathcal{U}(0.5,3.5)$  & 1.42$\pm$0.09 \\ \hline
\multicolumn{3}{c}{Epoch 1 \cht\ sine model parameters}\\ \hline
Semi-Amplitude $A$ ($\%$) & $\mathcal{U}(0,0.5)$  & 0.58$\pm$0.09 \\ 
Period $P$ (hr) & $\mathcal{U}(0,20)$  & 6.5$\pm$0.2 \\ 
Phase $\phi$ (degrees) & $\mathcal{U}(-90,450)$  & 269$\pm$31 \\ 
Constant $C$ & $\mathcal{U}(0.8,1.2)$  & 0.9994$\pm$0.0007 \\ 
Standard deviation $\sigma$ & $\mathcal{U}(0,1)$  & 0.0128$\pm$0.0005 \\ 
Bad data probability $P_{bad}$ & $\mathcal{U}(0,1)$  & $0.07^{+0.02}_{-0.01}$ \\ 
Bad data standard deviation $\sigma_{bad}$ & $\mathcal{U}(0,4)$  & $0.10^{+0.02}_{-0.01}$ \\ 
Bad data mean $Y_{bad}$ & $\mathcal{U}(0.5,3.5)$  & 1.04$\pm$0.02 \\ \hline
\multicolumn{3}{c}{Epoch 2 \cho\ constant model parameters}\\ \hline
Constant $C$ & $\mathcal{U}(C>0)$  & 0.997$\pm$0.002 \\ 
Standard deviation $\sigma$ & $\mathcal{U}(0,1)$  & 0.031$\pm$0.001 \\ 
Bad data probability $P_{bad}$ & $\mathcal{U}(0,1)$  & 0.05$\pm$0.01 \\ 
Bad data standard deviation $\sigma_{bad}$ & $\mathcal{U}(0,4)$  & $0.24^{+0.05}_{-0.03}$ \\ 
Bad data mean $Y_{bad}$ & $\mathcal{U}(0.5,3.5)$  & 1.24$\pm$0.06 \\ \hline 
\multicolumn{3}{c}{Epoch 2 \cht\ sine model parameters}\\ \hline
Semi-Amplitude $A$ ($\%$) & $\mathcal{U}(0,0.5)$  & 0.44$\pm$0.07 \\ 
Period $P$ (hr) & $\mathcal{U}(0,20)$  & $3.15^{+0.07}_{-0.06}$ \\ 
Phase $\phi$ (degrees) & $\mathcal{U}(-90,450)$  & $179^{+42}_{-44}$ \\ 
Constant $C$ & $\mathcal{U}(0.8,1.2)$  & 0.9996$\pm$0.0005 \\ 
Standard deviation $\sigma$ & $\mathcal{U}(0,1)$  & 0.0093$\pm$0.0003 \\ 
Bad data probability $P_{bad}$ & $\mathcal{U}(0,1)$  & 0.04$\pm$0.01 \\ 
Bad data standard deviation $\sigma_{bad}$ & $\mathcal{U}(0,4)$  & $0.12^{+0.03}_{-0.02}$ \\ 
Bad data mean $Y_{bad}$ & $\mathcal{U}(0.5,3.5)$  & 1.06$\pm$0.03 \\ 
\enddata
\tablenotetext{a}{$\mathcal{U}(a,b)$ denotes a uniform distribution over the range $a$ to $b$.}
\tablenotetext{b}{The values reported correspond to the 16th, 50th, and 84th percentiles of the marginalized posterior distribution.}
\end{deluxetable}
\newpage
%%%%%%%%%%%%%%%%%%%%%%%%%%%%%%%%%%%%%%%%%%%% W0734-7157 %%%%%%%%%%%%%%%%%%%%%%%%%%%%%%%%%%%%%%%%%%%%

\begin{deluxetable}{llc}[!ht]
\tablecaption{WISEJ0734$-$71 \label{tab:W0734}}
\tablecolumns{3}
\tablewidth{1.0\columnwidth}
\tablehead{
\colhead{Model Parameter} & \colhead{Prior\tablenotemark{a}} & \colhead{Value\tablenotemark{b}}
}
\startdata
\multicolumn{3}{c}{Epoch 1 \cho\ constant model parameters}\\ \hline
Constant $C$ & $\mathcal{U}(C>0)$  & 0.997$\pm$0.003 \\ 
Standard deviation $\sigma$ & $\mathcal{U}(0,1)$  & 0.056$\pm$0.002 \\ 
Bad data probability $P_{bad}$ & $\mathcal{U}(0,1)$  & 0.04$\pm$0.01 \\ 
Bad data standard deviation $\sigma_{bad}$ & $\mathcal{U}(0,4)$  & $0.44^{+0.10}_{-0.08}$ \\ 
Bad data mean $Y_{bad}$ & $\mathcal{U}(0.5,3.5)$  & 1.7$\pm$0.1 \\ \hline
\multicolumn{3}{c}{Epoch 1 \cht\ constant model parameters}\\ \hline
Constant $C$ & $\mathcal{U}(C>0)$  & 0.9997$\pm$0.0009 \\ 
Standard deviation $\sigma$ & $\mathcal{U}(0,1)$  & $0.0175^{+0.0008}_{-0.0007}$ \\ 
Bad data probability $P_{bad}$ & $\mathcal{U}(0,1)$  & $0.09^{+0.02}_{-0.01}$ \\ 
Bad data standard deviation $\sigma_{bad}$ & $\mathcal{U}(0,4)$  & 0.22$\pm$0.03 \\ 
Bad data mean $Y_{bad}$ & $\mathcal{U}(0.5,3.5)$  & 1.15$\pm$0.04 \\ \hline
\multicolumn{3}{c}{Epoch 2 \cho\ constant model parameters}\\ \hline
Constant $C$ & $\mathcal{U}(C>0)$  & 1.000$\pm$0.003 \\ 
Standard deviation $\sigma$ & $\mathcal{U}(0,1)$  & $0.057^{+0.003}_{-0.002}$ \\ 
Bad data probability $P_{bad}$ & $\mathcal{U}(0,1)$  & $0.06^{+0.02}_{-0.01}$ \\ 
Bad data standard deviation $\sigma_{bad}$ & $\mathcal{U}(0,4)$  & $0.40^{+0.07}_{-0.05}$ \\ 
Bad data mean $Y_{bad}$ & $\mathcal{U}(0.5,3.5)$  & $1.33^{+0.10}_{-0.09}$ \\ \hline
\multicolumn{3}{c}{Epoch 2 \cht\ constant model parameters}\\ \hline
Constant $C$ & $\mathcal{U}(C>0)$  & 0.9991$\pm$0.0009 \\ 
Standard deviation $\sigma$ & $\mathcal{U}(0,1)$  & 0.0171$\pm$0.0007 \\ 
Bad data probability $P_{bad}$ & $\mathcal{U}(0,1)$  & $0.09^{+0.02}_{-0.01}$ \\ 
Bad data standard deviation $\sigma_{bad}$ & $\mathcal{U}(0,4)$  & $0.19^{+0.03}_{-0.02}$ \\ 
Bad data mean $Y_{bad}$ & $\mathcal{U}(0.5,3.5)$  & 1.11$\pm$0.03 \\ \hline
\enddata
\tablenotetext{a}{$\mathcal{U}(a,b)$ denotes a uniform distribution over the range $a$ to $b$.}
\tablenotetext{b}{The values reported correspond to the 16th, 50th, and 84th percentiles of the marginalized posterior distribution.}
\end{deluxetable}
\newpage

%%%%%%%%%%%%%%%%%%%%%%%%%%%%%%%%%%%%%%%%%%%% WISE 0806B %%%%%%%%%%%%%%%%%%%%%%%%%%%%%%%%%%%%%%%%

\begin{deluxetable}{llc}[!ht]
\tablecaption{WD0806$-$661B\tablenotemark{a} \label{tab:W0806}}
\tablecolumns{3}
\tablewidth{1.0\columnwidth}
\tablehead{
\colhead{Model Parameter} & \colhead{Prior\tablenotemark{b}} & \colhead{Value\tablenotemark{c}}
}
\startdata
\multicolumn{3}{c}{Epoch 1 \cho\ constant model parameters}\\ \hline
Constant $C$ & $\mathcal{U}(C>0)$  & $0.975\pm 0.009$ \\
Standard deviation $\sigma$ & $\mathcal{U}(0,1)$  & 0.175 $\pm$ 0.007 \\
Bad data probability $P_{bad}$ & $\mathcal{U}(0,1)$  & 0.04 $\pm$ 0.01 \\
Bad data standard deviation $\sigma_{bad}$ & $\mathcal{U}(0,4)$  & $1.5^{+0.4}_{-0.3}$ \\
Bad data mean $Y_{bad}$ & $\mathcal{U}(0.5,3.5)$  & $2.3\pm 0.4$ \\ \hline
\multicolumn{3}{c}{Epoch 1 \cht\ sine model parameters}\\ \hline
Amplitude $A$ ($\%$) & $\mathcal{U}(0,0.5)$  & $2.74\pm 0.5$ \\
Period $P$ (hr) & $\mathcal{U}(0,20)$  & $5.7^{+0.3}_{-0.2}$ \\
Phase $\phi$ (degrees) & $\mathcal{U}(-90,450)$  & $21^{+19}_{-18}$ \\
Constant $C$ & $\mathcal{U}(0.8,1.2)$  & $0.999\pm 0.003$ \\
Standard deviation $\sigma$ & $\mathcal{U}(0,1)$  & $0.059\pm 0.003$ \\
Bad data probability $P_{bad}$ & $\mathcal{U}(0,1)$  & $0.06^{+0.02}_{-0.01}$ \\
Bad data standard deviation $\sigma_{bad}$ & $\mathcal{U}(0,4)$  & $0.55^{+0.11}_{-0.08}$ \\
Bad data mean $Y_{bad}$ & $\mathcal{U}(0.5,2.5)$  & 1.3 $\pm$ 0.1 \\
\multicolumn{3}{c}{Epoch 2 \cht\ constant model parameters}\\ \hline
Constant $C$ & $\mathcal{U}(C>0)$  & $0.993\pm 0.004$ \\
Standard deviation $\sigma$ & $\mathcal{U}(0,1)$  & 0.074 $\pm$ 0.003 \\
Bad data probability $P_{bad}$ & $\mathcal{U}(0,1)$  & 0.06 $\pm$ 0.01 \\
Bad data standard deviation $\sigma_{bad}$ & $\mathcal{U}(0,4)$  & $0.8^{+0.2}_{-0.1}$ \\
Bad data mean $Y_{bad}$ & $\mathcal{U}(0.5,3.5)$  & $1.5\pm 0.2$ \\ \hline 
\enddata
\tablenotetext{a}{There was no data taken in Epoch 2 \cho.}
\tablenotetext{b}{$\mathcal{U}(a,b)$ denotes a uniform distribution over the range $a$ to $b$.}
\tablenotetext{c}{The values reported correspond to the 16th, 50th, and 84th percentiles of the marginalized posterior distribution.}
\end{deluxetable}

%%%%%%%%%%%%%%%%%%%%%%%%%%%%%%%%%%%%%%%%%%%% WISE 0855 %%%%%%%%%%%%%%%%%%%%%%%%%%%%%%%%%%%%%%%%%%

\begin{deluxetable}{llc}[!ht]
\tablecaption{WISEJ0855$-$07 \label{tab:W0855}}
\tablecolumns{3}
\tablewidth{1.0\columnwidth}
\tablehead{
\colhead{Model Parameter} & \colhead{Prior\tablenotemark{a}} & \colhead{Value\tablenotemark{b}}
}
\startdata
\multicolumn{3}{c}{Epoch 1 \cho\ constant model parameters}\\ \hline
Constant $C$ & $\mathcal{U}(C>0)$  & $0.997\pm 0.003$ \\
Standard deviation $\sigma$ & $\mathcal{U}(0,1)$  & 0.055 $\pm$ 0.002 \\
Bad data probability $P_{bad}$ & $\mathcal{U}(0,1)$  & 0.04 $\pm$ 0.01 \\
Bad data standard deviation $\sigma_{bad}$ & $\mathcal{U}(0,4)$  & $0.8^{+0.2}_{-0.1}$ \\
Bad data mean $Y_{bad}$ & $\mathcal{U}(0.5,3.5)$  & $1.6\pm 0.2$ \\ \hline
\multicolumn{3}{c}{Epoch 1 \cht\ double sine model parameters}\\ \hline
Amplitude 1 $A_{1}$ ($\%$) & $\mathcal{U}(0,.5)$  & $1.49^{+0.10}_{-0.08}$ \\
Amplitude 2 $A_{2}$ ($\%$) & $\mathcal{U}(0,.5)$  & $0.94^{+0.08}_{-0.11}$ \\
Period $P$ (hr) & $\mathcal{U}(0,20)$  & $11.5^{+0.5}_{-0.7} $ \\
Phase 1 $\phi_{1}$ (degrees) & $\mathcal{U}(-90,450)$  & $ 331^{+5}_{-7}$ \\
Phase 2 $\phi_{2}$ (degrees) & $\mathcal{U}(-90,450)$  & $191^{+14}_{-20}$ \\
Constant $C$ & $\mathcal{U}(0.8,1.2)$  & 1.0013$\pm$ 0.0006 \\
Standard deviation $\sigma$ & $\mathcal{U}(0,1)$  & 0.0073 $\pm$ 0.0003 \\
Bad data probability $P_{bad}$ & $\mathcal{U}(0,1)$  & 0.04 $\pm$ 0.01 \\
Bad data standard deviation $\sigma_{bad}$ & $\mathcal{U}(0,4)$  & $0.12^{+0.03}_{-0.02}$ \\
Bad data mean $Y_{bad}$ & $\mathcal{U}(0.5,2.5)$  & 1.10 $\pm$ 0.03 \\
\multicolumn{3}{c}{Epoch 2 \cho\ constant model parameters}\\ \hline
Constant $C$ & $\mathcal{U}(C>0)$  & 0.993 $\pm$ 0.003 \\
Standard deviation $\sigma$ & $\mathcal{U}(0,1)$  & 0.058 $\pm$ 0.002 \\
Bad data probability $P_{bad}$ & $\mathcal{U}(0,1)$  & 0.03 $\pm$ 0.01 \\
Bad data standard deviation $\sigma_{bad}$ & $\mathcal{U}(0,4)$  & $0.7^{+0.2}_{-0.1}$ \\
Bad data mean $Y_{bad}$ & $\mathcal{U}(0.5,3.5)$  & $1.7\pm 0.2$ \\ \hline
\multicolumn{3}{c}{Epoch 2 \cht\ sine model parameters}\\ \hline
Amplitude $A$ ($\%$) & $\mathcal{U}(0,0.5)$  & $2.0^{+0.2}_{-0.1}$ \\
Period $P$ (hr) & $\mathcal{U}(0,20)$  & $17.0\pm 1$ \\
Phase $\phi$ (degrees) & $\mathcal{U}(-90,450)$  & $285^{+12}_{-10}$ \\
Constant $C$ & $\mathcal{U}(0.8,1.2)$  & $0.990^{+0.001}_{-0.002}$ \\
Standard deviation $\sigma$ & $\mathcal{U}(0,1)$  & $0.0085^{+0.0004}_{-0.0003}$ \\
Bad data probability $P_{bad}$ & $\mathcal{U}(0,1)$  & $0.08^{+0.02}_{-0.01}$ \\
Bad data standard deviation $\sigma_{bad}$ & $\mathcal{U}(0,4)$  & 0.08 $\pm$ 0.01 \\
Bad data mean $Y_{bad}$ & $\mathcal{U}(0.5,2.5)$  & 1.06 $\pm$ 0.02 \\
\enddata
\tablenotetext{a}{$\mathcal{U}(a,b)$ denotes a uniform distribution over the range $a$ to $b$.}
\tablenotetext{b}{The values reported correspond to the 16th, 50th, and 84th percentiles of the marginalized posterior distribution.}
\end{deluxetable}
\newpage

%%%%%%%%%%%%%%%%%%%%%%%%%%%%%%%%%%%%%%%%%%%% W1405+5534 %%%%%%%%%%%%%%%%%%%%%%%%%%%%%%%%%%%%%%%%%%%%

\begin{deluxetable}{llc}[!ht]
\tablecaption{WISEJ1405$+$55 \label{tab:W1405}}
\tablecolumns{3}
\tablewidth{1.0\columnwidth}
\tablehead{
\colhead{Model Parameter} & \colhead{Prior\tablenotemark{a}} & \colhead{Value\tablenotemark{b}}
}
\startdata
\multicolumn{3}{c}{Epoch 1 \cho\ constant model parameters}\\ \hline
Constant $C$ & $\mathcal{U}(C>0)$  & 0.998$\pm$0.002 \\ 
Standard deviation $\sigma$ & $\mathcal{U}(0,1)$  & $0.038^{+0.002}_{-0.001}$ \\ 
Bad data probability $P_{bad}$ & $\mathcal{U}(0,1)$  & 0.06$\pm$0.01 \\ 
Bad data standard deviation $\sigma_{bad}$ & $\mathcal{U}(0,4)$  & $0.36^{+0.06}_{-0.05}$ \\ 
Bad data mean $Y_{bad}$ & $\mathcal{U}(0.5,3.5)$  & $1.36^{+0.09}_{-0.08}$ \\ \hline
\multicolumn{3}{c}{Epoch 1 \cht\ double sine model parameters}\\ \hline
Semi-Amplitude 1 $A_{1}$ ($\%$) & $\mathcal{U}(0,.5)$  & 0.71$\pm$0.07 \\ 
Semi-Amplitude 2 $A_{2}$ ($\%$) & $\mathcal{U}(0,.5)$  & 0.58$\pm$0.07 \\ 
Period $P$ (hr) & $\mathcal{U}(0,20)$  & $9.1^{+0.4}_{-0.3}$ \\ 
Phase 1 $\phi_{1}$ (degrees) & $\mathcal{U}(-90,450)$  & $95^{+32}_{-28}$ \\ 
Phase 2 $\phi_{2}$ (degrees) & $\mathcal{U}(-90,450)$  & $349^{+58}_{-49}$ \\ 
Constant $C$ & $\mathcal{U}(0.8,1.2)$  & 1.0005$\pm$0.0005 \\ 
Standard deviation $\sigma$ & $\mathcal{U}(0,1)$  & 0.0092$\pm$0.0004 \\ 
Bad data probability $P_{bad}$ & $\mathcal{U}(0,1)$  & 0.05$\pm$0.01 \\ 
Bad data standard deviation $\sigma_{bad}$ & $\mathcal{U}(0,4)$  & $0.11^{+0.03}_{-0.02}$ \\ 
Bad data mean $Y_{bad}$ & $\mathcal{U}(0.5,3.5)$  & 1.06$\pm$0.03 \\ \hline
\multicolumn{3}{c}{Epoch 2 \cho\ sine model parameters}\\ \hline
Semi-Amplitude $A$ ($\%$) & $\mathcal{U}(0,0.5)$  & 3.7$\pm$0.3 \\ 
Period $P$ (hr) & $\mathcal{U}(0,20)$  & $8.4^{+0.3}_{-0.2}$ \\ 
Phase $\phi$ (degrees) & $\mathcal{U}(-90,450)$  & 201$\pm$9 \\ 
Constant $C$ & $\mathcal{U}(0.8,1.2)$  & 1.008$\pm$0.002 \\ 
Standard deviation $\sigma$ & $\mathcal{U}(0,1)$  & 0.036$\pm$0.001 \\ 
Bad data probability $P_{bad}$ & $\mathcal{U}(0,1)$  & 0.05$\pm$0.01 \\ 
Bad data standard deviation $\sigma_{bad}$ & $\mathcal{U}(0,4)$  & $0.42^{+0.08}_{-0.06}$ \\ 
Bad data mean $Y_{bad}$ & $\mathcal{U}(0.5,3.5)$  & 1.4$\pm$0.1 \\ \hline 
\multicolumn{3}{c}{Epoch 2 \cht\ double sine model parameters}\\ \hline
Semi-Amplitude 1 $A_{1}$ ($\%$) & $\mathcal{U}(0,0.5)$  & 3.51$\pm$0.07 \\ 
Semi-Amplitude 2 $A_{2}$ ($\%$) & $\mathcal{U}(0,0.5)$  & 0.47$\pm$0.07 \\ 
Period $P$ (hr) & $\mathcal{U}(0,20)$  & $8.42^{+0.06}_{-0.05}$ \\ 
Phase 1 $\phi_{1}$ (degrees) & $\mathcal{U}(-90,450)$  & 203$\pm$5 \\ 
Phase 2 $\phi_{2}$ (degrees) & $\mathcal{U}(-90,450)$  & 302$\pm$13 \\ 
Constant $C$ & $\mathcal{U}(0.8,1.2)$  & 0.9870$\pm$0.0005 \\ 
Standard deviation $\sigma$ & $\mathcal{U}(0,1)$  & 0.0094$\pm$0.0004 \\ 
Bad data probability $P_{bad}$ & $\mathcal{U}(0,1)$  & 0.06$\pm$0.01 \\ 
Bad data standard deviation $\sigma_{bad}$ & $\mathcal{U}(0,4)$  & $0.24^{+0.04}_{-0.03}$ \\ 
Bad data mean $Y_{bad}$ & $\mathcal{U}(0.5,3.5)$  & 1.10$\pm$0.05 \\ 
\enddata
\tablenotetext{a}{$\mathcal{U}(a,b)$ denotes a uniform distribution over the range $a$ to $b$.}
\tablenotetext{b}{The values reported correspond to the 16th, 50th, and 84th percentiles of the marginalized posterior distribution.}
\end{deluxetable}
\newpage
%%%%%%%%%%%%%%%%%%%%%%%%%%%%%%%%%%%%%%%%%%%% W1541-2250 %%%%%%%%%%%%%%%%%%%%%%%%%%%%%%%%%%%%%%%%%%%%

\begin{deluxetable}{llc}[!ht]
\tablecaption{WISEJ1541$-$22 \label{tab:W1541}}
\tablecolumns{3}
\tablewidth{1.0\columnwidth}
\tablehead{
\colhead{Model Parameter} & \colhead{Prior\tablenotemark{a}} & \colhead{Value\tablenotemark{b}}
}
\startdata
\multicolumn{3}{c}{Epoch 1 \cho\ constant model parameters}\\ \hline
Constant $C$ & $\mathcal{U}(C>0)$  & 0.998$\pm$0.002 \\ 
Standard deviation $\sigma$ & $\mathcal{U}(0,1)$  & 0.029$\pm$0.001 \\ 
Bad data probability $P_{bad}$ & $\mathcal{U}(0,1)$  & 0.04$\pm$0.01 \\ 
Bad data standard deviation $\sigma_{bad}$ & $\mathcal{U}(0,4)$  & $0.31^{+0.08}_{-0.06}$ \\ 
Bad data mean $Y_{bad}$ & $\mathcal{U}(0.5,3.5)$  & 1.33$\pm$0.09 \\ \hline
\multicolumn{3}{c}{Epoch 1 \cht\ constant model parameters}\\ \hline
Constant $C$ & $\mathcal{U}(C>0)$  & 0.9995$\pm$0.0005 \\ 
Standard deviation $\sigma$ & $\mathcal{U}(0,1)$  & 0.0107$\pm$0.0004 \\ 
Bad data probability $P_{bad}$ & $\mathcal{U}(0,1)$  & 0.05$\pm$0.01 \\ 
Bad data standard deviation $\sigma_{bad}$ & $\mathcal{U}(0,4)$  & 0.11$\pm$0.02 \\ 
Bad data mean $Y_{bad}$ & $\mathcal{U}(0.5,3.5)$  & 1.11$\pm$0.03 \\ \hline
\multicolumn{3}{c}{Epoch 2 \cho\ constant model parameters}\\ \hline
Constant $C$ & $\mathcal{U}(C>0)$  & $0.997^{+0.002}_{-0.001}$ \\ 
Standard deviation $\sigma$ & $\mathcal{U}(0,1)$  & 0.029$\pm$0.001 \\ 
Bad data probability $P_{bad}$ & $\mathcal{U}(0,1)$  & 0.05$\pm$0.01 \\ 
Bad data standard deviation $\sigma_{bad}$ & $\mathcal{U}(0,4)$  & $0.23^{+0.05}_{-0.04}$ \\ 
Bad data mean $Y_{bad}$ & $\mathcal{U}(0.5,3.5)$  & 1.30$\pm$0.06 \\ \hline
\multicolumn{3}{c}{Epoch 2 \cht\ constant model parameters}\\ \hline
Constant $C$ & $\mathcal{U}(C>0)$  & 0.9997$\pm$0.0005 \\ 
Standard deviation $\sigma$ & $\mathcal{U}(0,1)$  & $0.0098^{+0.0004}_{-0.0003}$ \\ 
Bad data probability $P_{bad}$ & $\mathcal{U}(0,1)$  & 0.04$\pm$0.01 \\ 
Bad data standard deviation $\sigma_{bad}$ & $\mathcal{U}(0,4)$  & $0.09^{+0.02}_{-0.01}$ \\ 
Bad data mean $Y_{bad}$ & $\mathcal{U}(0.5,3.5)$  & 1.09$\pm$0.02 \\ \hline
\enddata
\tablenotetext{a}{$\mathcal{U}(a,b)$ denotes a uniform distribution over the range $a$ to $b$.}
\tablenotetext{b}{The values reported correspond to the 16th, 50th, and 84th percentiles of the marginalized posterior distribution.}
\end{deluxetable}
\newpage
%%%%%%%%%%%%%%%%%%%%%%%%%%%%%%%%%%%%%%%%%%%% W1639-6847 %%%%%%%%%%%%%%%%%%%%%%%%%%%%%%%%%%%%%%%%%%%%

\begin{deluxetable}{llc}[!ht]
\tablecaption{WISEJ1639$-$68 \label{tab:W1639}}
\tablecolumns{3}
\tablewidth{1.0\columnwidth}
\tablehead{
\colhead{Model Parameter} & \colhead{Prior\tablenotemark{a}} & \colhead{Value\tablenotemark{b}}
}
\startdata
\multicolumn{3}{c}{Epoch 1 \cho\ constant model parameters}\\ \hline
Constant $C$ & $\mathcal{U}(C>0)$  & 1.003$\pm$0.002 \\ 
Standard deviation $\sigma$ & $\mathcal{U}(0,1)$  & 0.030$\pm$0.001 \\ 
Bad data probability $P_{bad}$ & $\mathcal{U}(0,1)$  & 0.02$\pm$0.01 \\ 
Bad data standard deviation $\sigma_{bad}$ & $\mathcal{U}(0,4)$  & $0.21^{+0.08}_{-0.05}$ \\ 
Bad data mean $Y_{bad}$ & $\mathcal{U}(0.5,3.5)$  & $1.21^{+0.09}_{-0.08}$ \\ \hline
\multicolumn{3}{c}{Epoch 1 \cht\ double sine model parameters}\\ \hline
Semi-Amplitude 1 $A_{1}$ ($\%$) & $\mathcal{U}(0,.5)$  & 0.42$\pm$0.05 \\ 
Semi-Amplitude 2 $A_{2}$ ($\%$) & $\mathcal{U}(0,.5)$  & 0.38$\pm$0.05 \\ 
Period $P$ (hr) & $\mathcal{U}(0,20)$  & $6.32^{+0.09}_{-0.08}$ \\ 
Phase 1 $\phi_{1}$ (degrees) & $\mathcal{U}(-90,450)$  & $53^{+16}_{-15}$ \\ 
Phase 2 $\phi_{2}$ (degrees) & $\mathcal{U}(-90,450)$  & $73^{+30}_{-28}$ \\ 
Constant $C$ & $\mathcal{U}(0.8,1.2)$  & 0.9994$\pm$0.0003 \\ 
Standard deviation $\sigma$ & $\mathcal{U}(0,1)$  & $0.0067^{+0.0003}_{-0.0002}$ \\ 
Bad data probability $P_{bad}$ & $\mathcal{U}(0,1)$  & 0.02$\pm$0.01 \\ 
Bad data standard deviation $\sigma_{bad}$ & $\mathcal{U}(0,4)$  & $0.11^{+0.04}_{-0.03}$ \\ 
Bad data mean $Y_{bad}$ & $\mathcal{U}(0.5,3.5)$  & 1.07$\pm$0.04 \\ \hline
\multicolumn{3}{c}{Epoch 2 \cho\ sine model parameters}\\ \hline
Semi-Amplitude $A$ ($\%$) & $\mathcal{U}(0,0.5)$  & 1.6$\pm$0.2 \\ 
Period $P$ (hr) & $\mathcal{U}(0,20)$  & 6.8$\pm$0.2 \\ 
Phase $\phi$ (degrees) & $\mathcal{U}(-90,450)$  & 44$\pm$11 \\ 
Constant $C$ & $\mathcal{U}(0.8,1.2)$  & 0.999$\pm$0.001 \\ 
Standard deviation $\sigma$ & $\mathcal{U}(0,1)$  & 0.0222$\pm$0.0009 \\ 
Bad data probability $P_{bad}$ & $\mathcal{U}(0,1)$  & 0.07$\pm$0.01 \\ 
Bad data standard deviation $\sigma_{bad}$ & $\mathcal{U}(0,4)$  & $0.26^{+0.04}_{-0.03}$ \\ 
Bad data mean $Y_{bad}$ & $\mathcal{U}(0.5,3.5)$  & 1.19$\pm$0.05 \\ \hline 
\multicolumn{3}{c}{Epoch 2 \cht\ double sine model parameters}\\ \hline
Semi-Amplitude 1 $A_{1}$ ($\%$) & $\mathcal{U}(0,.5)$  & 1.52$\pm$0.05 \\ 
Semi-Amplitude 2 $A_{2}$ ($\%$) & $\mathcal{U}(0,.5)$  & 0.62$\pm$0.05 \\ 
Period $P$ (hr) & $\mathcal{U}(0,20)$  & 6.69$\pm$0.05 \\ 
Phase 1 $\phi_{1}$ (degrees) & $\mathcal{U}(-90,450)$  & 37$\pm$7 \\ 
Phase 2 $\phi_{2}$ (degrees) & $\mathcal{U}(-90,450)$  & $16^{+14}_{-15}$ \\ 
Constant $C$ & $\mathcal{U}(0.8,1.2)$  & 0.9992$\pm$0.0004 \\ 
Standard deviation $\sigma$ & $\mathcal{U}(0,1)$  & 0.0067$\pm$0.0003 \\ 
Bad data probability $P_{bad}$ & $\mathcal{U}(0,1)$  & $0.08^{+0.02}_{-0.01}$ \\ 
Bad data standard deviation $\sigma_{bad}$ & $\mathcal{U}(0,4)$  & $0.12^{+0.02}_{-0.01}$ \\ 
Bad data mean $Y_{bad}$ & $\mathcal{U}(0.5,3.5)$  & 1.07$\pm$0.02 \\ 
\enddata
\tablenotetext{a}{$\mathcal{U}(a,b)$ denotes a uniform distribution over the range $a$ to $b$.}
\tablenotetext{b}{The values reported correspond to the 16th, 50th, and 84th percentiles of the marginalized posterior distribution.}
\end{deluxetable}
\newpage
%%%%%%%%%%%%%%%%%%%%%%%%%%%%%%%%%%%%%%%%%%%% W1738+2732 %%%%%%%%%%%%%%%%%%%%%%%%%%%%%%%%%%%%%%%%%%%%

\begin{deluxetable}{llc}[!ht]
\tablecaption{WISEJ1738$+$27 \label{tab:W1738}}
\tablecolumns{3}
\tablewidth{1.0\columnwidth}
\tablehead{
\colhead{Model Parameter} & \colhead{Prior\tablenotemark{a}} & \colhead{Value\tablenotemark{b}}
}
\startdata
\multicolumn{3}{c}{Epoch 1 \cho\ sine model parameters}\\ \hline
Semi-Amplitude $A$ ($\%$) & $\mathcal{U}(0,0.5)$  & 1.7$\pm$0.3 \\ 
Period $P$ (hr) & $\mathcal{U}(0,20)$  & 4.8$\pm$0.2 \\ 
Phase $\phi$ (degrees) & $\mathcal{U}(-90,450)$  & 42$\pm$21 \\ 
Constant $C$ & $\mathcal{U}(0.8,1.2)$  & 0.996$\pm$0.002 \\ 
Standard deviation $\sigma$ & $\mathcal{U}(0,1)$  & 0.045$\pm$0.002 \\ 
Bad data probability $P_{bad}$ & $\mathcal{U}(0,1)$  & 0.05$\pm$0.01 \\ 
Bad data standard deviation $\sigma_{bad}$ & $\mathcal{U}(0,4)$  & $0.43^{+0.08}_{-0.06}$ \\ 
Bad data mean $Y_{bad}$ & $\mathcal{U}(0.5,3.5)$  & 1.5$\pm$0.1 \\ \hline 
\multicolumn{3}{c}{Epoch 1 \cht\ sine model parameters}\\ \hline
Semi-Amplitude $A$ ($\%$) & $\mathcal{U}(0,0.5)$  & 1.30$\pm$0.09 \\ 
Period $P$ (hr) & $\mathcal{U}(0,20)$  & $5.91^{+0.10}_{-0.09}$ \\ 
Phase $\phi$ (degrees) & $\mathcal{U}(-90,450)$  & 135$\pm$18 \\ 
Constant $C$ & $\mathcal{U}(0.8,1.2)$  & 0.9997$\pm$0.0005 \\ 
Standard deviation $\sigma$ & $\mathcal{U}(0,1)$  & 0.0117$\pm$0.0005 \\ 
Bad data probability $P_{bad}$ & $\mathcal{U}(0,1)$  & 0.05$\pm$0.01 \\ 
Bad data standard deviation $\sigma_{bad}$ & $\mathcal{U}(0,4)$  & 0.12$\pm$0.02 \\ 
Bad data mean $Y_{bad}$ & $\mathcal{U}(0.5,3.5)$  & 1.12$\pm$0.03 \\ \hline 
\multicolumn{3}{c}{Epoch 2 \cho\ constant model parameters}\\ \hline
Constant $C$ & $\mathcal{U}(C>0)$  & 0.998$\pm$0.002 \\ 
Standard deviation $\sigma$ & $\mathcal{U}(0,1)$  & 0.043$\pm$0.002 \\ 
Bad data probability $P_{bad}$ & $\mathcal{U}(0,1)$  & 0.04$\pm$0.01 \\ 
Bad data standard deviation $\sigma_{bad}$ & $\mathcal{U}(0,4)$  & $0.44^{+0.09}_{-0.07}$ \\ 
Bad data mean $Y_{bad}$ & $\mathcal{U}(0.5,3.5)$  & 1.4$\pm$0.1 \\ \hline
\multicolumn{3}{c}{Epoch 2 \cht\ double sine model parameters}\\ \hline
Semi-Amplitude 1 $A_{1}$ ($\%$) & $\mathcal{U}(0,.5)$  & 0.62$\pm$0.08 \\ 
Semi-Amplitude 2 $A_{2}$ ($\%$) & $\mathcal{U}(0,.5)$  & 0.66$\pm$0.08 \\ 
Period $P$ (hr) & $\mathcal{U}(0,20)$  & 6.09$\pm$0.09 \\ 
Phase 1 $\phi_{1}$ (degrees) & $\mathcal{U}(-90,450)$  & 216$\pm$19 \\ 
Phase 2 $\phi_{2}$ (degrees) & $\mathcal{U}(-90,450)$  & $12^{+32}_{-34}$ \\ 
Constant $C$ & $\mathcal{U}(0.8,1.2)$  & 0.9989$\pm$0.0006 \\ 
Standard deviation $\sigma$ & $\mathcal{U}(0,1)$  & 0.0107$\pm$0.0004 \\ 
Bad data probability $P_{bad}$ & $\mathcal{U}(0,1)$  & $0.08^{+0.02}_{-0.01}$ \\ 
Bad data standard deviation $\sigma_{bad}$ & $\mathcal{U}(0,4)$  & $0.12^{+0.02}_{-0.01}$ \\ 
Bad data mean $Y_{bad}$ & $\mathcal{U}(0.5,3.5)$  & 1.08$\pm$0.02 \\ 
\enddata
\tablenotetext{a}{$\mathcal{U}(a,b)$ denotes a uniform distribution over the range $a$ to $b$.}
\tablenotetext{b}{The values reported correspond to the 16th, 50th, and 84th percentiles of the marginalized posterior distribution.}
\end{deluxetable}
\newpage
%%%%%%%%%%%%%%%%%%%%%%%%%%%%%%%%%%%%%%%%%%%% W1828+2650 %%%%%%%%%%%%%%%%%%%%%%%%%%%%%%%%%%%%%%%%%%%%

\begin{deluxetable}{llc}[!ht]
\tablecaption{WISEJ1828$+$26 \label{tab:W1828}}
\tablecolumns{3}
\tablewidth{1.0\columnwidth}
\tablehead{
\colhead{Model Parameter} & \colhead{Prior\tablenotemark{a}} & \colhead{Value\tablenotemark{b}}
}
\startdata
\multicolumn{3}{c}{Epoch 1 \cho\ constant model parameters}\\ \hline
Constant $C$ & $\mathcal{U}(C>0)$  & 1.001$\pm$0.002 \\ 
Standard deviation $\sigma$ & $\mathcal{U}(0,1)$  & 0.040$\pm$0.02 \\ 
Bad data probability $P_{bad}$ & $\mathcal{U}(0,1)$  & 0.04$\pm$0.01 \\ 
Bad data standard deviation $\sigma_{bad}$ & $\mathcal{U}(0,4)$  & $0.6^{+0.2}_{-0.1}$ \\ 
Bad data mean $Y_{bad}$ & $\mathcal{U}(0.5,3.5)$  & 1.6$\pm$0.2 \\ \hline
\multicolumn{3}{c}{Epoch 1 \cht\ constant model parameters}\\ \hline
Constant $C$ & $\mathcal{U}(C>0)$  & 0.9992$\pm$0.0005 \\ 
Standard deviation $\sigma$ & $\mathcal{U}(0,1)$  & 0.0092$\pm$0.0003 \\ 
Bad data probability $P_{bad}$ & $\mathcal{U}(0,1)$  & 0.05$\pm$0.01 \\ 
Bad data standard deviation $\sigma_{bad}$ & $\mathcal{U}(0,4)$  & 0.07$\pm$0.01 \\ 
Bad data mean $Y_{bad}$ & $\mathcal{U}(0.5,3.5)$  & 1.07$\pm$0.02 \\ \hline
\multicolumn{3}{c}{Epoch 2 \cho\ constant model parameters}\\ \hline
Constant $C$ & $\mathcal{U}(C>0)$  & 0.999$\pm$0.002 \\ 
Standard deviation $\sigma$ & $\mathcal{U}(0,1)$  & 0.044$\pm$0.002 \\ 
Bad data probability $P_{bad}$ & $\mathcal{U}(0,1)$  & 0.05$\pm$0.01 \\ 
Bad data standard deviation $\sigma_{bad}$ & $\mathcal{U}(0,4)$  & 1.0$\pm$0.2 \\ 
Bad data mean $Y_{bad}$ & $\mathcal{U}(0.5,3.5)$  & 1.2$\pm$0.2 \\ \hline
\multicolumn{3}{c}{Epoch 2 \cht\ constant model parameters}\\ \hline
Constant $C$ & $\mathcal{U}(C>0)$  & 0.9996$\pm$0.0005 \\ 
Standard deviation $\sigma$ & $\mathcal{U}(0,1)$  & 0.0102$\pm$0.0004 \\ 
Bad data probability $P_{bad}$ & $\mathcal{U}(0,1)$  & $0.07^{+0.02}_{-0.01}$ \\ 
Bad data standard deviation $\sigma_{bad}$ & $\mathcal{U}(0,4)$  & $0.09^{+0.02}_{-0.01}$ \\ 
Bad data mean $Y_{bad}$ & $\mathcal{U}(0.5,3.5)$  & 1.08$\pm$0.02 \\ \hline
\enddata
\tablenotetext{a}{$\mathcal{U}(a,b)$ denotes a uniform distribution over the range $a$ to $b$.}
\tablenotetext{b}{The values reported correspond to the 16th, 50th, and 84th percentiles of the marginalized posterior distribution.}
\end{deluxetable}
\newpage
%%%%%%%%%%%%%%%%%%%%%%%%%%%%%%%%%%%%%%%%%%%% W2056+1459 %%%%%%%%%%%%%%%%%%%%%%%%%%%%%%%%%%%%%%%%%%%%

\begin{deluxetable}{llc}[!ht]
\tablecaption{WISEJ2056$+$14 \label{tab:W2056}}
\tablecolumns{3}
\tablewidth{1.0\columnwidth}
\tablehead{
\colhead{Model Parameter} & \colhead{Prior\tablenotemark{a}} & \colhead{Value\tablenotemark{b}}
}
\startdata
\multicolumn{3}{c}{Epoch 1 \cho\ constant model parameters}\\ \hline
Constant $C$ & $\mathcal{U}(C>0)$  & 0.998$\pm$0.001 \\ 
Standard deviation $\sigma$ & $\mathcal{U}(0,1)$  & 0.0220$\pm$0.0009 \\ 
Bad data probability $P_{bad}$ & $\mathcal{U}(0,1)$  & 0.06$\pm$0.01 \\ 
Bad data standard deviation $\sigma_{bad}$ & $\mathcal{U}(0,4)$  & 0.18$\pm$0.03 \\ 
Bad data mean $Y_{bad}$ & $\mathcal{U}(0.5,3.5)$  & $1.16^{+0.05}_{-0.04}$ \\ \hline
\multicolumn{3}{c}{Epoch 1 \cht\ constant model parameters}\\ \hline
Constant $C$ & $\mathcal{U}(C>0)$  & 0.9997$\pm$0.0004 \\ 
Standard deviation $\sigma$ & $\mathcal{U}(0,1)$  & 0.0082$\pm$0.0003 \\ 
Bad data probability $P_{bad}$ & $\mathcal{U}(0,1)$  & 0.04$\pm$0.01 \\ 
Bad data standard deviation $\sigma_{bad}$ & $\mathcal{U}(0,4)$  & 0.07$\pm$0.01 \\ 
Bad data mean $Y_{bad}$ & $\mathcal{U}(0.5,3.5)$  & 1.09$\pm$0.02 \\ \hline
\multicolumn{3}{c}{Epoch 2 \cho\ constant model parameters}\\ \hline
Constant $C$ & $\mathcal{U}(C>0)$  & 1.000$\pm$0.001 \\ 
Standard deviation $\sigma$ & $\mathcal{U}(0,1)$  & 0.0226$\pm$0.0009 \\ 
Bad data probability $P_{bad}$ & $\mathcal{U}(0,1)$  & 0.05$\pm$0.01 \\ 
Bad data standard deviation $\sigma_{bad}$ & $\mathcal{U}(0,4)$  & $0.23^{+0.04}_{-0.03}$ \\ 
Bad data mean $Y_{bad}$ & $\mathcal{U}(0.5,3.5)$  & $1.23^{+0.06}_{-0.05}$ \\ \hline
\multicolumn{3}{c}{Epoch 2 \cht\ constant model parameters}\\ \hline
Constant $C$ & $\mathcal{U}(C>0)$  & 1.0000$\pm$0.0004 \\ 
Standard deviation $\sigma$ & $\mathcal{U}(0,1)$  & $0.0080^{+0.0004}_{-0.0003}$ \\ 
Bad data probability $P_{bad}$ & $\mathcal{U}(0,1)$  & $0.08^{+0.02}_{-0.01}$ \\ 
Bad data standard deviation $\sigma_{bad}$ & $\mathcal{U}(0,4)$  & 0.08$\pm$0.01 \\ 
Bad data mean $Y_{bad}$ & $\mathcal{U}(0.5,3.5)$  & 1.05$\pm$0.02 \\ \hline
\enddata
\tablenotetext{a}{$\mathcal{U}(a,b)$ denotes a uniform distribution over the range $a$ to $b$.}
\tablenotetext{b}{The values reported correspond to the 16th, 50th, and 84th percentiles of the marginalized posterior distribution.}
\end{deluxetable}
\newpage
%%%%%%%%%%%%%%%%%%%%%%%%%%%%%%%%%%%%%%%%%%%% W2220-3628 %%%%%%%%%%%%%%%%%%%%%%%%%%%%%%%%%%%%%%%%%%%%

\begin{deluxetable}{llc}[!ht]
\tablecaption{WISEJ2220$-$36 \label{tab:W2220}}
\tablecolumns{3}
\tablewidth{1.0\columnwidth}
\tablehead{
\colhead{Model Parameter} & \colhead{Prior\tablenotemark{a}} & \colhead{Value\tablenotemark{b}}
}
\startdata
\multicolumn{3}{c}{Epoch 1 \cho\ constant model parameters}\\ \hline
Constant $C$ & $\mathcal{U}(C>0)$  & 1.001$\pm$0.002 \\ 
Standard deviation $\sigma$ & $\mathcal{U}(0,1)$  & 0.048$\pm$0.002 \\ 
Bad data probability $P_{bad}$ & $\mathcal{U}(0,1)$  & 0.04$\pm$0.01 \\ 
Bad data standard deviation $\sigma_{bad}$ & $\mathcal{U}(0,4)$  & $0.50^{+0.12}_{-0.08}$ \\ 
Bad data mean $Y_{bad}$ & $\mathcal{U}(0.5,3.5)$  & 1.2$\pm$0.1 \\ \hline
\multicolumn{3}{c}{Epoch 1 \cht\ constant model parameters}\\ \hline
Constant $C$ & $\mathcal{U}(C>0)$  & 0.9981$\pm$0.0007 \\ 
Standard deviation $\sigma$ & $\mathcal{U}(0,1)$  & 0.0133$\pm$0.0005 \\ 
Bad data probability $P_{bad}$ & $\mathcal{U}(0,1)$  & 0.07$\pm$0.01 \\ 
Bad data standard deviation $\sigma_{bad}$ & $\mathcal{U}(0,4)$  & 0.12$\pm$0.02 \\ 
Bad data mean $Y_{bad}$ & $\mathcal{U}(0.5,3.5)$  & $1.10^{+0.03}_{-0.02}$ \\ \hline
\multicolumn{3}{c}{Epoch 2 \cho\ constant model parameters}\\ \hline
Constant $C$ & $\mathcal{U}(C>0)$  & 0.997$\pm$0.003 \\ 
Standard deviation $\sigma$ & $\mathcal{U}(0,1)$  & 0.053$\pm$0.002 \\ 
Bad data probability $P_{bad}$ & $\mathcal{U}(0,1)$  & 0.05$\pm$0.01 \\ 
Bad data standard deviation $\sigma_{bad}$ & $\mathcal{U}(0,4)$  & $0.34^{+0.07}_{-0.06}$ \\ 
Bad data mean $Y_{bad}$ & $\mathcal{U}(0.5,3.5)$  & 1.4$\pm$0.1 \\ \hline
\multicolumn{3}{c}{Epoch 2 \cht\ sine model parameters}\\ \hline
Semi-Amplitude $A$ ($\%$) & $\mathcal{U}(0,0.5)$  & 0.7$\pm$0.1 \\ 
Period $P$ (hr) & $\mathcal{U}(0,20)$  & 3.23$\pm$0.08 \\ 
Phase $\phi$ (degrees) & $\mathcal{U}(-90,450)$  & 306$\pm$48 \\ 
Constant $C$ & $\mathcal{U}(0.8,1.2)$  & 1.0004$\pm$0.0008 \\ 
Standard deviation $\sigma$ & $\mathcal{U}(0,1)$  & 0.0158$\pm$0.0006 \\ 
Bad data probability $P_{bad}$ & $\mathcal{U}(0,1)$  & 0.04$\pm$0.01 \\ 
Bad data standard deviation $\sigma_{bad}$ & $\mathcal{U}(0,4)$  & $0.18^{+0.04}_{-0.03}$ \\ 
Bad data mean $Y_{bad}$ & $\mathcal{U}(0.5,3.5)$  & 1.14$\pm$0.05 \\ 
\enddata
\tablenotetext{a}{$\mathcal{U}(a,b)$ denotes a uniform distribution over the range $a$ to $b$.}
\tablenotetext{b}{The values reported correspond to the 16th, 50th, and 84th percentiles of the marginalized posterior distribution.}
\end{deluxetable}

\end{document}